\documentclass[12pt,a4paper]{article}
\usepackage{jheppub}
\pdfoutput=1
\usepackage{dsfont}
\usepackage{amssymb,amsmath}
\usepackage{epsfig}
\usepackage{epstopdf}
\usepackage{latexsym}
\usepackage{graphicx}
\usepackage{subfigure}
\usepackage{booktabs}
\usepackage{bbm}
\makeatletter
\def\@fpheader{\relax}
\makeatother

\def\p{\partial}

\newcommand{\be}{\begin{equation}}
\newcommand{\ee}{\end{equation}}
\newcommand{\bea}{\begin{eqnarray}}
\newcommand{\eea}{\end{eqnarray}}

\newcommand{\cO}{{\mathcal{O}}}
\newcommand{\cN}{{\mathcal{N}}}

\let\ev=\bracket

\subheader{\begin{flushright}
UTTG-09-14\\
TCC-009-14
\end{flushright}}

\title{Fluctuation and dissipation in de Sitter space}
\author[a,b]{Willy Fischler,}
\author[b,c]{Phuc H. Nguyen,}
\author[a,b]{Juan F. Pedraza}
\author[a,b]{and Walter Tangarife}
\affiliation[a]{Theory Group, Department of Physics, The University of Texas, Austin, TX 78712}
\affiliation[b]{Texas Cosmology Center, The University of Texas, Austin, TX 78712}
\affiliation[c]{Center for Relativity, Department of Physics, The University of Texas, Austin, TX 78712}
\emailAdd{fischler@physics.utexas.edu}
\emailAdd{phn229@physics.utexas.edu}
\emailAdd{jpedraza@physics.utexas.edu}
\emailAdd{wtang@physics.utexas.edu}

\abstract{In this paper we study some thermal properties of quantum field theories in de Sitter space by means of holographic techniques. We focus on the static patch of de Sitter and assume that the quantum fields are in the standard Bunch-Davies vacuum. More specifically, we follow the stochastic motion of a massive charged particle due to its interaction with Hawking radiation. The process is described in terms of the theory of Brownian motion in inhomogeneous media and its associated Langevin dynamics. At late times, we find that the particle undergoes a regime of slow diffusion and never reaches the horizon, in stark contrast to the usual random walk behavior at finite temperature. Nevertheless, the fluctuation-dissipation theorem is found to hold at all times.}

\begin{document}

\maketitle
\flushbottom

\section{Introduction}

Quantum field theory in curved space is a subject of great
interest that has led to many interesting areas of research in the past few decades \cite{bd,Wald:1995yp}.
In this context, gravity itself is treated classically but matter fields can interact quantum mechanically.
Although it is believed to be a good physical description in many circumstances, it is clear that a fully quantum theory of gravitational interactions is needed to address problems where effective QFT breaks down, leading to various puzzles
and paradoxes \cite{Giddings:2006sj,Mathur:2009hf,Banks:2010tj,Almheiri:2012rt,braunstein}. Nevertheless, this framework has provided us with a useful tool to investigate certain aspects in quantum gravity and has led, for instance, to a better understanding of black hole thermodynamics and the physical consequences of inflationary cosmology.

QFT in curved space is often discussed in terms of free or weakly coupled field theories, in which case perturbative calculations are under control. While most of the qualitative features are visible at this level, understanding the strong-coupling and non-perturbative regimes is also of great interest. In particular, in the context of cosmology it is plausible that strong interactions might have played an important role in the early universe, creating a demand for new theoretical tools. In recent years, significant steps have been taken towards meeting this challenge via the dS/CFT \cite{Strominger:2001pn,McInnes:2001zw,Strominger:2001gp} and AdS/CFT correspondences \cite{Maldacena:1997re,Gubser:1998bc,Witten:1998qj}, as well as other holographic scenarios \cite{Banks:1996vh,Fischler:1998st,Alishahiha:2004md,Freivogel:2006xu,Sekino:2009kv,Dong:2010pm,Banks:2011av,Banks:2011qf,Banks:2013fr,Banks:2013fri}.

We focus on cosmological models proposed in the context of AdS/CFT or, more generally, the gauge/gravity correspondence (see \cite{Buchel:2002wf,Buchel:2002kj,Aharony:2002cx,Balasubramanian:2002am,Cai:2002mr,Silverstein:2003hf,Ross:2004cb,Balasubramanian:2005bg,Alishahiha:2005dj,Freivogel:2005qh,Buchel:2006em,Hirayama:2006jn,Ghoroku:2006af,He:2007ji,Ghoroku:2006nh,Hutasoit:2009xy,Marolf:2010tg,Li:2011bt,Erdmenger:2011sy,Ghoroku:2012vi,Buchel:2013dla,Fischler:2013fba} for references).
Among them, there are a number of proposals for theories living in a de Sitter universe. We will follow closely the work of \cite{Marolf:2010tg}. In that paper, the authors considered CFTs with bulk duals coming from the standard Einstein-Hilbert action with negative cosmological constant. These can be thought of as models that belong to a universality class of strongly-coupled gauge theories, in the large-$N$ limit. In the bulk, the solutions are described by a class of AdS black hole solutions with special topology, the so-called hyperbolic black holes \cite{Emparan:1998he,Birmingham:1998nr,Emparan:1999gf}. In particular, the zero mass limit of these black holes was identified as the dual of the CFT in the Bunch-Davies vacuum state. More in general, these solutions correspond to different states of the CFT that are kept in thermal equilibrium (in the static patch of de Sitter), with an extra term in the holographic stress-energy tensor that can be identified as a radiation source. More recently, the authors of \cite{Fischler:2013fba} have also considered out-of-equilibrium situations interpolating between two of the above states, by generalizing the bulk solutions to the well-known Vaidya geometries.

Our goal is to study thermal properties of these theories, focusing for simplicity on the Bunch-Davies vacuum state. In this state, a static observer equipped with an Unruh-DeWitt detector will see a thermal bath of Hawking quanta at a temperature proportional to the Hubble constant \cite{Gibbons:1977mu,Bunch:1978yq},\footnote{From here on, we will work in units such that $\hbar=k_B=1$. However, we should keep in mind that thermal effects in de Sitter space are of quantum mechanical nature.}
\be
T_{\text{dS}}=\frac{ H  }{2\pi } \frac{\hbar}{k_B}\ .
\ee
Some preliminary results were obtained in \cite{Hutasoit:2009xy,Fischler:2013fba}. In particular, the holographic computation of entanglement entropy in the Bunch-Davies vacuum \cite{Fischler:2013fba} shows an interesting behavior: while it gives the expected result due to the gravitational conformal anomaly in curved space, it fails to give the appropriate scaling characteristic of QFT at finite temperature.\footnote{In $3+1$ dimensions it was found that, for a spherical region of radius $R<1/H$, the correction to the entanglement entropy due to the curvature of the space scales like $S_H(R)-S_0(R)\sim H^2R^2\sim T_{\text{dS}}^2R^2$. However, in quantum field theory at finite temperature the expected behavior is $S_T(R)-S_0(R)\sim T^4R^4$. On the other hand, for odd number of dimensions one finds that $S_H(R)-S_0(R)=0$, which is consistent with the fact that in such cases there is no gravitational conformal anomaly. We believe that one of the reasons that accounts for this particular behavior is the fact that entanglement entropy is a static observable. In the present paper we will focus on other thermal properties that can probe the real-time dynamics of the theories.} Given this antecedent, a natural question that arises in this context is to what extent the thermal behaviour of these de Sitter QFTs agrees (or not) with the expectations from finite-temperature field theories.

This paper is devoted to further our understanding of real-time dynamics in de Sitter space when the system is in the Bunch-Davies vacuum state.
In this line of research, one could compute the energy-momentum correlators, \emph{e.g.} $\langle T_{\mu\nu}(x)T_{\alpha\beta}(y)\rangle$, and determine from these the various transport coefficients of the plasma. In principle, this could be done by following the steps of \cite{Hutasoit:2009xy}, which studied scalar and current correlators in the same hyperbolic black hole geometries. However, in the static patch of de Sitter there is no invariance under spatial translations, adding some mathematical (and technical) complications that are not easy to overcome. Instead, we will consider external probes, and analyze the manner in which the plasma damps their motion. More specifically, we will consider the dynamics of a heavy quark that is introduced externally and interacts quantum mechanically with the bath of Hawking radiation.

In the holographic context, a heavy quark on the boundary theory corresponds to the
endpoint of an open string that stretches between the boundary and
the black hole horizon \cite{Rey:1998ik}. In this picture, the quark is described in a first-quantized language and it couples to the strongly-coupled
field-theoretic degrees of freedom, whose path integral is fully carried out via bulk dynamics. The seminal works \cite{Herzog:2006gh,Gubser:2006bz} focused on the energy loss of a quark
that is either moving with constant velocity, or is moving non-relativistically and about to come to rest.
In these two setups the string is treated classically, implying that the path integral over
the quark trajectory is treated in a saddle-point approximation. On the other hand, in order to study the fluctuations of the quark due to its interaction with the thermal bath one must go beyond the classical description of the string. As customary, fluctuations of the string around an average embedding are described
in terms of free scalar fields propagating on the induced worldsheet
geometry. These fields can then be excited quantum mechanically due to Hawking radiation, populating the various
modes of oscillation of the string. Finally, the induced motion of the string endpoint is found to be described in terms of Brownian motion and its associated Langevin dynamics \cite{brownian,Son:2009vu}.\footnote{Some earlier studies include \cite{gubserqhat,ctqhat}.}$^{,}$\footnote{At weak coupling, some aspects of Brownian motion in de Sitter space and Rindler space have been studied in \cite{Hu:1993qa,Jaekel:1997hr,Fingerle:2007tn,Haba:2009by,Iso:2010yq,Kolekar:2012sf,Rigopoulos:2013exa,Oshita:2014dha}. The latter case is of particular relevance in the near-horizon region of de Sitter, and have been considered in the context of holography in \cite{Caceres:2010rm}.} These results were later elaborated on in \cite{Caceres:2010rm,Giecold:2009cg,CasalderreySolana:2009rm,Atmaja:2010uu,Das:2010yw,Gursoy:2010aa,Ebrahim:2010ra,CaronHuot:2011dr,Kiritsis:2011bw,
Fischler:2012ff,Tong:2012nf,Edalati:2012tc,Atmaja:2012jg,Sadeghi:2013lka,Atmaja:2013gxa,Banerjee:2013rca,Giataganas:2013hwa,Yeh:2013mca,
Kiritsis:2013iba,Chakrabortty:2013kra,Sadeghi:2013jja,Giataganas:2013zaa,Sadeghi:2014lha}.

The rest of the paper is organized as follows. In section \ref{sec2}, we introduce the gravity duals of de Sitter QFTs studied in \cite{Marolf:2010tg,Fischler:2013fba}. In section \ref{sec3} we briefly review the dynamics of a Brownian particle in a thermal bath. We start in \ref{ldsec} by presenting some of the most important features of the Langevin equation. Next, in \ref{dssec}, we move on to describe new features that arise in the description of Brownian motion in de Sitter space and we propose a suitable generalization of
the Langevin equation that captures these effects.
In section \ref{hbmsec} we show how to realize this phenomenon at strong-coupling, in terms of a probe string living in the bulk geometry.
In \ref{subsub1} we study the dynamics of the string in the bulk, setting the grounds of our holographic computations. In \ref{subsub2} and \ref{admittanceSEC} we compute two-point functions and the response function, respectively. This allows us to extract the random force correlator and the admittance, which characterize the Langevin equation of the heavy quark. Finally, in \ref{remarks}, we comment on the lessons learned from the holographic computation with respect to the phenomenological equation proposed previously in \ref{dssec}. In section \ref{sec5}, we give a summary of the main results and close with conclusions. We relegate some of the more technical computations to the appendices.

\section{Holographic models of de Sitter QFTs\label{sec2}}

The purpose of this section is to give a quick review of the gravity duals of de Sitter QFTs first studied in \cite{Marolf:2010tg} and elaborated on in \cite{Fischler:2013fba}.\footnote{For a recent review of QFT on curved space via holography see \cite{Marolf:2013ioa}.} The starting point is the $(d+1)$-dimensional Einstein-Hilbert action with negative cosmological constant,
\be\label{action0}
S = \frac{1}{2\kappa^2} \int d^{d+1} x \sqrt{-g} \left(R - 2 \Lambda \right)\,.
\ee
From this action we obtain the following equations of motion
\be\label{Eineom}
R_{\mu\nu} - \frac{1}{2} \left(R- 2 \Lambda \right) g_{\mu\nu}=0\,,
\ee
where $\kappa^2=8\pi G_N^{(d+1)}$ and $\Lambda=-d(d-1)/2L^2$.

The idea is to find solutions to (\ref{Eineom}) with a foliation such that the boundary metric corresponds to some parametrization of de Sitter space. This, by itself, does not uniquely determine the bulk metric. To see this more directly, let us write the metric in the Fefferman-Graham form \cite{fg}
\begin{equation}\label{feffermangraham}
ds^2={L^2 \over z^2}\left(g_{\mu\nu}(z,x)dx^{\mu}dx^{\nu}+dz^2\right)~.
\end{equation}
The metric of the boundary CFT can be directly read off as $g_{\mu\nu}(x)= g_{\mu\nu}(0,x)$. Additionally, the full function $g_{\mu\nu}(z,x)$ encodes data dual to the expectation value of the boundary stress-energy tensor $T_{\mu\nu}(x)$. In terms of the near-boundary expansion
\begin{equation}\label{metricexpansion}
g_{\mu\nu}(z,x)=g_{\mu\nu}(x)+z^2 g^{(2)}_{\mu\nu}(x)+\ldots +z^d g^{(d)}_{\mu\nu}(x)+ z^d \log (z^2) h^{(d)}_{\mu\nu}(x)+\ldots~,
\end{equation}
the standard GKPW recipe for correlation functions \cite{Gubser:1998bc,Witten:1998qj} leads, after holographic renormalization, to \cite{dhss,skenderis,skenderis2}
\begin{equation}\label{graltmunu}
\left\langle T_{\mu\nu}(x)\right\rangle={ d\, L^{d-1} \over 16\pi G^{(d+1)}_{N}}\left(g^{(d)}_{\mu\nu}(x)+X^{(d)}_{\mu\nu}(x)\right)~,
\end{equation}
where $X^{(d)}_{\mu\nu}=0$ $\forall$ odd $d$,
\begin{eqnarray}\label{xmunu}
X^{(2)}_{\mu\nu}&=&-g_{\mu\nu}g^{(2)\alpha}_{\alpha}~,\\
X^{(4)}_{\mu\nu}&=&-{1\over 8}g_{\mu\nu}\left[\left(g_{\alpha}^{(2)\alpha}\right)^2
-g_{\alpha}^{(2)\beta}g_{\beta}^{(2)\alpha}\right]
-{1\over 2}g_{\mu}^{(2)\alpha}g_{\alpha\nu}^{(2)}
+{1\over 4}g^{(2)}_{\mu\nu}g_{\alpha}^{(2)\alpha}~,\nonumber
\end{eqnarray}
and $X^{(2d)}_{\mu\nu}$ for $d\geq3$ given by similar but longer expressions that we will not transcribe here. Thus, it is clear that in order to find a gravity solution we must specify both, a boundary metric and the state of the dual theory, which are encoded in the non-normalizable and normalizable modes of the bulk geometry.
\begin{figure}[t!]
\begin{center}
  \includegraphics[width=7cm]{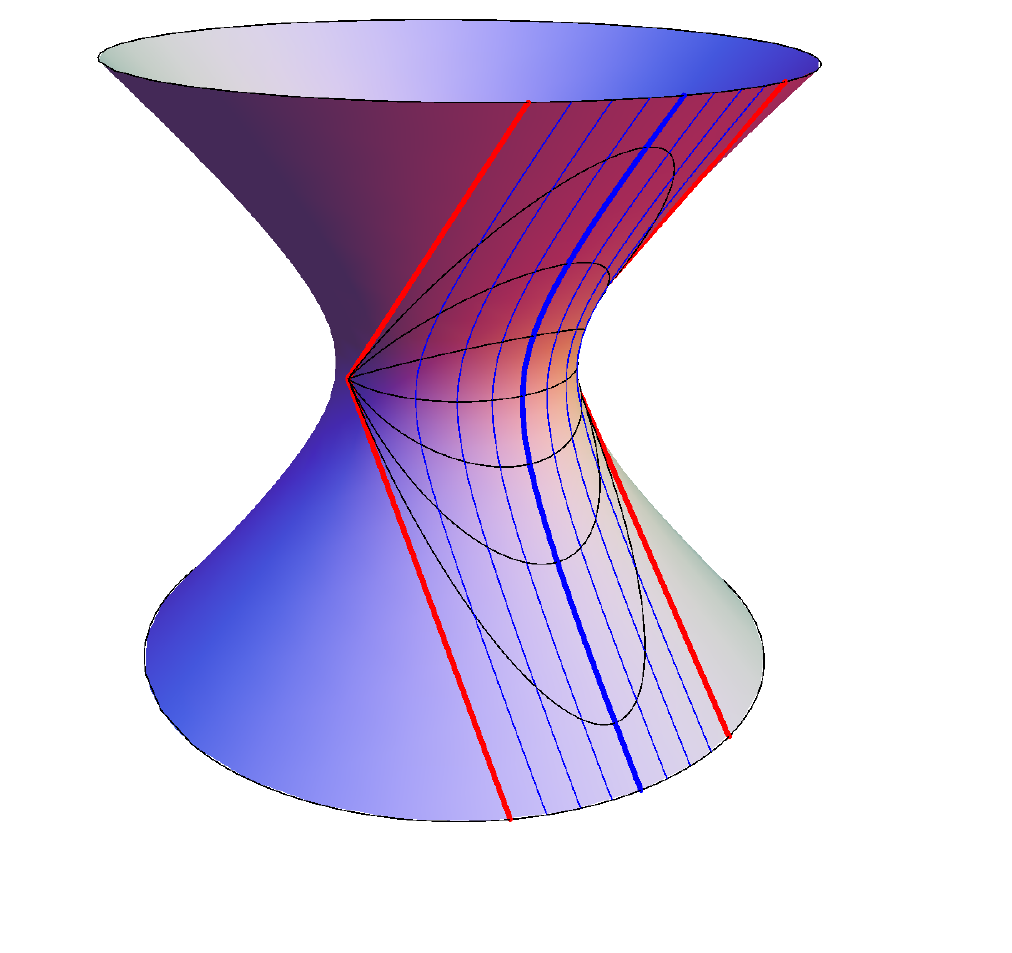}
\end{center}
\vspace{-1.4cm}
\caption{The static chart on the de Sitter manifold, represented as a hyperboloid embedded in Minkowski space. The thick blue curve represents the worldline of the static observer. The light blue curves represent observers at constant $r$. Such observers are not in free fall; they can be thought of as traveling on rockets toward the geodesic observer, but due to the expansion of space, actually never meet the observer at $r=0$. The horizon corresponds to the red curves, and the black curves are constant $t$ slices.\label{static}}
\end{figure}

For free (or weakly coupled) field theories in de Sitter space, there is a family of de Sitter invariant vacuum states known as the $\alpha$-vacua \cite{Mottola:1984ar,Allen:1985ux}. However, among these states only the Bunch-Davies (or Euclidean) vacuum \cite{Bunch:1978yq} reduces to the standard Minkowski vacuum in the limit $H\to0$. We are mainly interested in this state for the purposes of the present paper. The Bunch-Davies vacuum is well defined on the entire manifold but, for concreteness, we will focus in the static patch of de Sitter, which covers the causal diamond associated with a single geodesic observer,
\be\label{staticdS}
ds^2=-(1-H^2r^2)dt^2+\frac{dr^2}{1-H^2r^2}+r^2d\Omega^2_{d-2}\,,
\ee
where $H$ denotes the value of the Hubble constant. In Figure \ref{static} we show a diagram of the static patch of de Sitter space. The name ``static'' comes from the fact that there is a killing vector $\xi=\partial_t$ associated with the isometry of time translations.\footnote{In particular, this implies that correlators of the form $\langle\phi(t_0)\phi(t_0+t)\rangle$ are independent of $t_0$.} Therefore, energy as well as entropy are well defined quantities. For the static observer, the Bunch-Davies vacuum is characterized by a temperature that is associated to the presence of a cosmological horizon \cite{Gibbons:1977mu}. More specifically, the observer will see a horizon at $r=1/H$ with surface gravity
\begin{equation}\label{surfacegravity}
   k_h \equiv -{1 \over 2}\left(\nabla_{\mu} \xi_{\nu} \nabla^{\mu} \xi^{\nu}\right)_{\scriptsize
   \mbox{horizon}}=H~,
\end{equation}
which enters a classical ``first law of event horizons'' reminiscent of black hole thermodynamics. This similarity is shown to be more than an analogy: if the observer is equipped with a particle detector, it will indeed observe a background of Hawking radiation at a temperature $T\equiv k_h/2\pi=T_{\text{dS}}$, coming from the cosmological event horizon.

Gravitational solutions to (\ref{Eineom}) dual to strongly-coupled theories in de Sitter space were given in \cite{Marolf:2010tg,Fischler:2013fba}. In particular, the bulk metric corresponding to the Bunch-Davies vacuum was identified as\footnote{In string theory constructions, this metric will be multiplied by a compact manifold that will not play any role in our discussion.}
\be
ds^2= \frac{L^2}{z^2}\left[\left(1-\frac{H^2z^2}{4}\right)^2\left(-(1-H^2r^2)dt^2+\frac{dr^2}{1-H^2r^2} + r^2 d\Omega^2_{d-2}\right)+dz^2 \right]. \label{metric1}
\ee
This geometry is completely smooth and absent of singularities. The solution has a regular Killing horizon at $z=2/H$ in addition to the expected cosmological horizon at $r=1/H$ $\forall$ $z$, both of which with temperature $T=T_{\text{dS}}$.

\begin{figure}[t!]
\begin{center}
  \includegraphics[width=5cm]{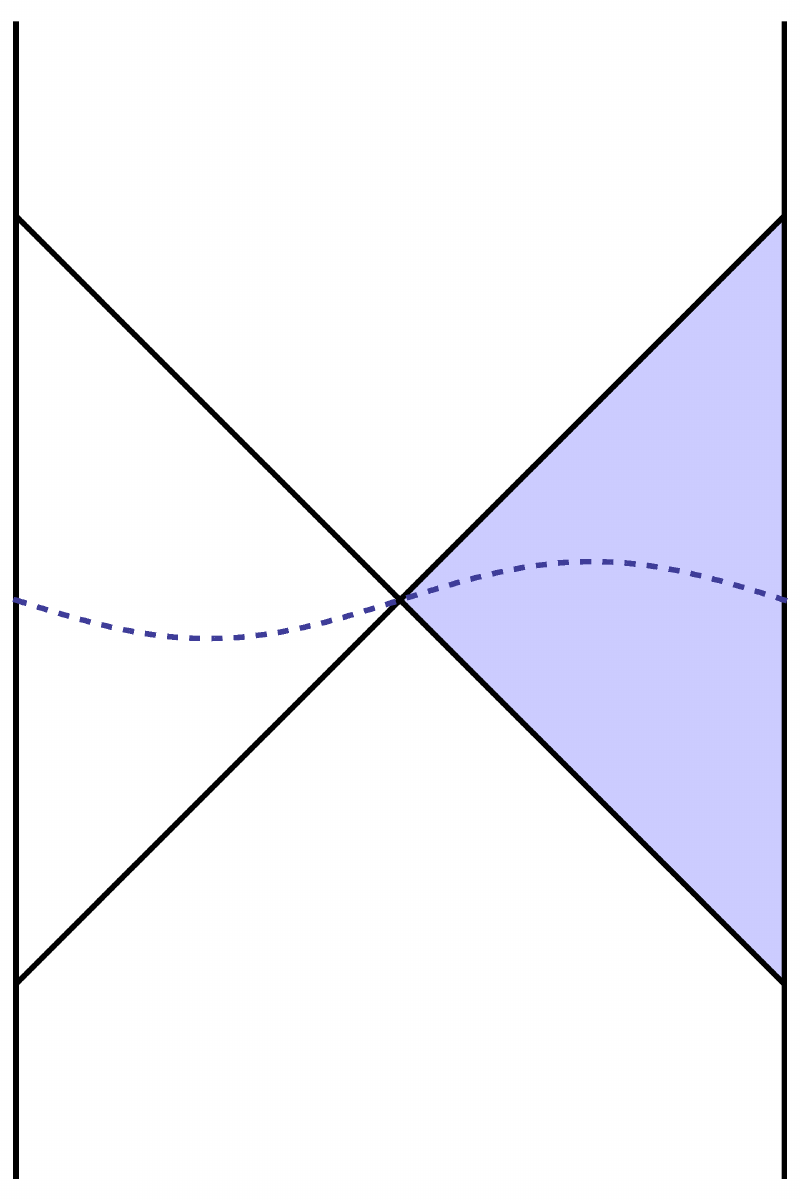}
\end{center}
\begin{picture}(0,0)
 \put(295,133){\small  $z=0$}
 \put(225,95){\small  $z=\tfrac{2}{H}$}
  \put(225,170){\small  $z=\tfrac{2}{H}$}
 \end{picture}
\vspace{-0.8cm}
\caption{Region of AdS covered by the hyperbolic patch. The Killing horizon in this case is analogous to a Rindler horizon, with an associated temperature and non-vanishing area. The dashed line in this diagram correspond to a Cauchy surface.\label{hyperbolicpatch}}
\end{figure}
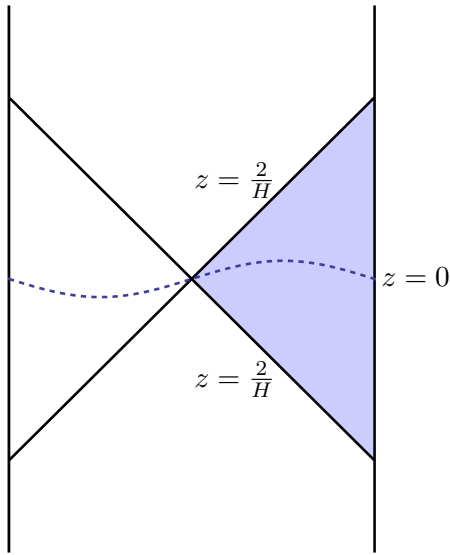

It can be shown that, under an appropriate bulk diffeomorphism, the metric (\ref{metric1}) is related to the zero mass limit of the so-called hyperbolic (or topological) black holes described in \cite{Emparan:1998he,Birmingham:1998nr,Emparan:1999gf}.\footnote{As mentioned in the introduction, the authors of \cite{Marolf:2010tg} used these hyperbolic black holes to construct the gravity dual of de Sitter QFTs in more general states (in thermal equilibrium). More recently, time dependent configurations were studied in \cite{Fischler:2013fba} by considering the Vaidya generalization of the same black hole geometries.} For zero mass, the solution is isometric to AdS and is completely non-singular. However, it covers a smaller portion of the entire manifold, which is sometimes referred to as the hyperbolic patch of AdS--- see Figure \ref{hyperbolicpatch}. The Killing horizon in this case is analogous to a Rindler horizon, with an associated temperature and non-vanishing area.

\section{Review on Brownian motion\label{sec3}}

In this section we will revisit the effective field theory description of Brownian motion in a thermal plasma. Along the way, we will stress on some phenomenological properties of this phenomenon in de Sitter space and we will propose a suitable generalization of the Langevin equation that captures the new effects arising in this geometry.

\subsection{Langevin dynamics in a nutshell\label{ldsec}}

In 1827, Robert Brown noticed that pollen particles suspended in a fluid undergo an erratic motion \cite{Brown}, which is known nowadays as Brownian motion. Mathematically, this process is described in terms of a Langevin equation which takes into account the random kicks received by the constituents of the fluid \cite{Uhlenbeck:1930zz,Chandrasekhar:1943ws,UhlenbeckII}. In its simplest version, the Langevin equation describing the motion of a massive particle in one
spatial dimension can be written as
\be
 \dot p(t)=-\gamma_0 \,p(t)+R(t) \ ,
\label{simpleLE}
\ee
where $p(t)=m\,v(t)$ is the momentum of the particle, $m$ its mass and $v(t)=\dot{x}(t)$ its velocity. In equation (\ref{simpleLE}), we assumed that the particle is much heavier than the fluid constituents and, therefore, its motion is non-relativisitic. The terms appearing in the right-hand side of this equation are the friction and random forces, respectively, and the constant $\gamma_0$ is known as the friction coefficient. In this description, the particle loses energy to the thermal bath because of the friction term and, at the same time, receives random kicks as prescribed by the random force.

As a first approximation, we can model the random force as white noise, with the following statistical averages:
\be\label{rforce}
\langle R(t)\rangle=0,\qquad\langle R(t)R(t')\rangle=\kappa_0\delta(t-t') \ ,
\ee
where the constant $\kappa_0$ measures the strength of the random force. Due to the fluctuation-dissipation theorem, this constant is related to the friction coefficient through
\begin{align}\label{flucdis}
 \gamma_0&={\kappa_0\over 2\,m\,T} \ .
\end{align}
This results from the fact that friction and random forces originate from the same mechanism at the microscopic level, \emph{i.e.} collisions with the particles of the thermal bath.

One can explore the physical consequences of equation (\ref{simpleLE}). One important result concerns to the time evolution of the
mean squared displacement \cite{Uhlenbeck:1930zz}:
\begin{align}
 \ev{s(t)^2}\equiv
 \ev{[x(t)-x(0)]^2}
 ={2\,T\over\gamma_0^2 \,m}\, \left(\gamma_0 \,t-1+e^{-\gamma_0 \,t}\right)
 \approx
 \begin{cases}
  \displaystyle {T\over m}\,t^2 & \displaystyle \left(\gamma_0\,t\ll 1\right) \ ,\\[3ex]
  \displaystyle \frac{2\,T}{\gamma_0 \,m}\,t           & \displaystyle \left(\gamma_0\,t\gg 1\right) \ .
 \end{cases}
 \label{s^2_simple}
\end{align}
For $t\ll\gamma_0^{-1}$, the motion is inertial ($s\sim v\, t$) with a velocity given by the equipartition of energy, $m\,v^2\sim T$. On the other hand, for $t\gg\gamma_0^{-1}$, the particle behaves as in a random walk ($s\sim \sqrt{D\,t}$). This crossover between the \emph{ballistic} and \emph{diffusive} regimes takes place in a time scale of order
\be
 t_{\text{relax}}\sim \gamma_0^{-1},
\label{t_relax}
\ee
in which the Brownian particle loses memory of its initial conditions and thermalizes.

In $n>1$ spatial dimensions we can write down an equation equivalent to (\ref{simpleLE}), where $p$ and $R$ are
replaced by $n$-component vectors, and $\gamma_0$ and $\kappa_0$ become rank-two tensors, $\Gamma$ and $\mathcal{K}$ respectively. If the plasma is isotropic, \emph{i.e.} if $\Gamma_{ij}=\gamma_0\,\delta_{ij}$ and $\mathcal{K}_{ij}=\kappa_0\, \delta_{ij}$ for $i,j=1,\dots, n$, then the equations of motion along the different directions decouple.
In such case, the displacement squared (\ref{s^2_simple}) at late times
goes instead as $\ev{s(t)^2}\approx 2nT t/\gamma_0 m$. The fluctuation-dissipation theorem (\ref{flucdis}), on the other hand, is
unaffected by the number of dimensions.

\subsection{General considerations in de Sitter space\label{dssec}}

Now let us focus on the de Sitter space. In this case, there are at least two new effects that cannot be accounted for in the usual Langevin equation (\ref{simpleLE}). The first one is that, from the point of view of a static observer (sitting at $r=0$), any particle away from the origin will feel a gravitational force acting upon it. For non-relativistic motion this force is simply given by
\be\label{gravforce}
\vec{F}_{\text{grav}}=-m \vec{\nabla} \Phi,
\ee
where $\Phi$ denotes the Newtonian gravitational potential. This can be directly read off from the 00-component of the metric tensor, which in the weak field limit can be written as $g_{00}=-(1+2\Phi)$.\footnote{See for instance \cite{WB}.} In particular, for the de Sitter metric (\ref{staticdS}) we obtain
\be\label{pot}
\Phi(r)=-\frac{1}{2}H^2r^2 \ ,
\ee
\emph{i.e.} a quadratic inverted potential. This means that any massive particle away from the origin will feel a repulsive force, $\vec{F}_{\text{dS}}=m H^2r\hat{r}$, that pushes it towards the horizon.

The second effect is the redshift of energies induced by the gravitational field. More specifically,
the way a static observer has access to other points of spacetime and in particular, to our Brownian particle, is by collecting `photons' or 'gravitons' that are radiated away (or reflected) by the particle itself. Due to the gravitational redshift, the wavelength
or energy observed at the center is actually lower than the energy of the particle when it was created. This is ultimately a result of the gravitational time dilation--- the fact that the rate at which time passes depends on a local clock at each point of spacetime.

These two effects, the gravitational force and the redshift of energies, are correctly encoded in the classical equation of motion for a test particle--- see appendix \ref{geoapp} for details. In particular, radial time-like geodesics in de Sitter space satisfy
\be\label{radgeo}
\ddot{r}(t)=H^2 r(t)\left(1-H^2 r(t)^2\right)-\frac{3 H^2 r(t) \dot{r}(t)^2}{1-H^2 r(t)^2}\ ,
\ee
where $\dot{\,}\equiv d/dt$. In addition to this, the particle can interact quantum mechanically with the bath of Hawking radiation emitted by the cosmological horizon, effect that is not encoded in (\ref{radgeo}). Phenomenologically, this could be taken into account
by putting this particle at finite temperature in a similar way to equation (\ref{simpleLE}). However, given the structure of (\ref{radgeo}), one can see that this would lead to a nonlinear Langevin equation. The problem becomes analytically tractable by focusing on the regime $\dot{r}\ll1$ and $Hr\ll1$. In this case, (\ref{radgeo}) leads to
\be
\ddot{r}(t)=H^2 r(t) \ ,
\ee
which is linear in $r(t)$. Notice that this is exactly the equation of motion of a particle in an inverted quadratic potential of the form (\ref{pot}), and is valid for non-relativistic motion and small deviations from the origin. On the other hand, in de Sitter space the physics is isotropic as seen by a static observer. Therefore, we can focus on the case of Brownian motion along one of the spatial dimensions ($n=1$), say $x$.
In the presence of a thermal bath, one could postulate the following generalization of the Langevin equation (\ref{simpleLE}):\footnote{In the non-relativistic limit $p(t)=m\dot{x}(t)$.}
\be\label{LangeHP}
\dot{p}(t)=-\gamma_0 \,p(t)+F_{\text{grav}}(t)+R(t) \ ,
\ee
with $F_{\text{grav}}(t)=m H^2 x(t)$ and $R(t)$ satisfying (\ref{rforce}). Equation (\ref{LangeHP}) should capture
the main properties of the physics, but it fails to give a physically consistent picture for sufficiently short times $t$, in which the particle suffers only a few or no impacts. It is a general feature of any dynamical system that the dynamical coherence becomes predominant in short time scales, or at high frequencies. In this regime, equation (\ref{LangeHP}) is not realistic enough for the following reasons: first,
it assumes that the
friction acts instantaneously and, second, that the
random force is uncorrelated at different times. In a general QFT, the interactions between the Brownian particle and the medium are nonlinear and these two assumptions will no longer hold;
the friction will depend on the trajectory of the particle, and random forces at different
times will be generally correlated. In de Sitter space, we expect these two effects to receive further corrections coming from the curvature of spacetime. In particular, it is known that for orbits of the killing field $\xi$ away from the origin, the temperature will experience a gravitational blueshift according to \cite{bd,Wald:1995yp}
\begin{equation}\label{localtemp}
   T(r)={k_h\over 2\pi \sqrt{-\xi\cdot\xi}}
   =\frac{T_{\text{dS}}}{\sqrt{1-H^2r^2}}=T_{\text{dS}}\left(1+\frac{1}{2}H^2r^2+\mathcal{O}(H^4r^4)\right)~.
\end{equation}
This means that the Brownian particle will have to interact with more and more Hawking quanta as it moves away from the origin. This new feature in de Sitter space is expected to modify substantially the physics at late times, or equivalently, low frequencies. Nevertheless, we will assume that the combination of all these effects can be taken into account via a generalized Langevin equation \cite{Mori:genLE,Kubo:f-d_thm},
\begin{gather}
\dot{p}(t)=-\int_{-\infty}^t dt'\, \gamma(t-t')\, p(t')-k\, x(t)+R(t)+F_{\text{ext}}(t)\, , \qquad k\equiv -m\, H^2 \,,
 \label{genLE}
\end{gather}
where the friction is now nonlocal and depends on the entire history of the motion through the memory
kernel $\gamma(t)$. The random force is also generalized to
\begin{align}
 \ev{R(t)}=0,\qquad
 \ev{R(t)R(t')}=\kappa(t-t'),
 \label{RRcorr}
\end{align}
for some function $\kappa(t)$. Moreover, we have included an extra term $F_{\text{ext}}(t)$ which is an external
force that can be applied to the particle.

We believe that some of these new effects in de Sitter space can be inferred from the theory of Brownian motion in an inhomogeneous medium, a topic that has been studied extensively in the condensed matter literature \cite{BrwonianI1,BrwonianI2,BrwonianI3,BrwonianI4,BrwonianI5,BrwonianI6,BrwonianI7,BrwonianI8,BrwonianI9} and is known to have some peculiar properties.  In the presence of a potential field, for instance, a hot zone drives the diffusing particles up the potential slope. This is the so-called blow-torch theorem \cite{BrwonianI9}. Diffusion properties in inhomogeneous media have no universal generalization, and the behavior depends on the precise physical mechanism that causes the diffusion \cite{BrwonianI7}. In general, the late time behaviour of the displacement squared is modified according to
\be
 \ev{s(t)^2}\sim D\, t^\alpha\,,\qquad\gamma_0\,t\gg 1 \ ,
\ee
where $D$ is the diffusion constant. If $\alpha>1$, the phenomenon is called super-diffusion. The case $0\leq \alpha < 1$, on the other hand, is known as sub-diffusion. Recently, this kind of anomalous behavior was observed experimentally in several systems including ultra-cold atoms \cite{Anomalous1}, and various biological systems \cite{Anomalous2,Anomalous3,Anomalous4}.

One way to analyze the diffusion properties of an inhomogeneous system is by considering a phenomenological equation that takes into account an effective force due to the gradient of temperature \cite{BrwonianI8}
\be
F_{\text{eff}}=-\frac{d \Phi}{d x}+a(T)\frac{d T}{d x} \ ,
\ee
where $a(T)$ is a function that depends on the microscopic model of the host medium. If the force is confining (\emph{i.e.} if it diverges for $x\to\infty$) equipartition does not generally hold and the stationary solution of the diffusion equation will \emph{not} be a Maxwell-Boltzmann distribution. In such cases the particle is expected to undergo sub-diffusion $0\leq \alpha < 1$. We argue that Brownian motion in de Sitter space corresponds to the critically damped case $\alpha=0$ for one simple reason: the classical equation of motion for a radial geodesic (\ref{radgeo}) predicts itself that $r(t)\to1/H$ as $t\to\infty$ and the inclusion of temperature cannot make it better. We conclude that, in this case, it is reasonable to expect that $\ev{s(t)^2}\to D$ as $t\to\infty$, where $D<1/H^2$, \emph{i.e.} that there is an equilibrium point at which the gravitational repulsion balances with the effects of the gradient of temperature. We will corroborate this by explicit computation in section \ref{subsub2}. In addition, we will show in appendix \ref{subsub4} that the distribution of velocities does not follow a Maxwell-Boltzmann distribution.

Before closing this section, let us analyze some properties of the generalized Langevin equation (\ref{genLE}). In frequency domain, this equation can be written as\footnote{Causality imposes that $\gamma(t)=0$ for $t<0$ so $\gamma[\omega]$ in this expression denotes the Fourier-Laplace transform,
$$\gamma[\omega]=\int_{0}^\infty dt\, \gamma(t) \,e^{i\omega t},$$
while $p(\omega)$, $R(\omega)$, and $F_{\text{ext}}(\omega)$ are Fourier transforms, e.g.,
$$p(\omega)=\int_{-\infty}^\infty dt\, p(t)\,e^{i\omega t}.$$}
\begin{equation}\label{genLEomega}
p(\omega) = \frac{R(\omega)+F_{\text{ext}}(\omega)}{\gamma[\omega]-i\omega+i\frac{k}{\omega m}}\,.
\end{equation}
Taking the statistical average of the same, we get
\begin{equation}\label{admitt}
\langle p(\omega) \rangle = \mu{(\omega)}F_{\text{ext}}{(\omega)}\,,
\end{equation}
where the quantity
\begin{equation}\label{mu1}
\mu(\omega) = \frac{1}{\gamma[\omega]-i\omega+i\frac{k}{\omega m}}
\end{equation}
is known as the admittance. Thus, we can determine the
admittance $\mu(\omega)$, and hence $\gamma[\omega]$, by applying an external force $F_{\text{ext}}{(\omega)}$ to the Brownian particle and then
measuring the response $\left\langle p(\omega)\right\rangle$.  In
particular, if we take
\be
 F_{\text{ext}}(t)=F_0e^{-i\omega t}\,,
\ee
then $\left\langle p(t)\right\rangle$ evaluates to
\be
 \left\langle p(t)\right\rangle=\mu(\omega) F_0e^{-i\omega t}=\mu(\omega) F_{\text{ext}}(t)\,.
\ee
From the admittance $\mu(\omega)$ we can extract
\begin{equation}\label{frictioncoefff}
\gamma[\omega] = \frac{1}{\mu{(\omega)}}+i\omega-i\frac{k}{\omega m}\,,
\end{equation}
and, in the low frequency limit,
\begin{equation}\label{frictioncoeff}
\gamma_0 = \lim_{\omega \rightarrow 0} \mathbf{Re}{\left(\frac{1}{\mu{(\omega)}}\right)}\,,\qquad \frac{k}{m} = \lim_{\omega \rightarrow 0} \mathbf{Im}{\left(\frac{\omega}{\mu{(\omega)}}\right)}\,.
\end{equation}
Finally, when the external force is set to zero in (\ref{genLEomega}), it follows that
\be
 p(\omega)={R(\omega)\over \gamma[\omega]-i\omega+i\frac{k}{\omega m}}=\mu(\omega)R(\omega),
\ee
and then, the power spectrum reads\footnote{For a quantity $\cO$, the power spectrum $I_\cO(\omega)$ is defined as
$$
  I_\cO(\omega)\equiv\int_{-\infty}^\infty dt\left\langle\cO(t)\cO(0)\right\rangle e^{i\omega t},
 \label{pwrspctr_def}
$$
and it is related to the two-point function through the Wiener-Khintchine theorem
$$
\left\langle\cO(\omega)\cO(\omega')\right\rangle=2\pi \delta(\omega+\omega')I_\cO(\omega).
$$}
\be
 I_{p}(\omega)=\left|\mu(\omega)\right|^2I_{R}(\omega).
\ee
Therefore, the random force correlator appearing in (\ref{RRcorr}) can be obtained from
\be\label{kapa}
\kappa(\omega)=I_{R}(\omega)=\frac{I_{p}(\omega)}{\left|\mu(\omega)\right|^2}\,,
\ee
and, in the low frequency limit,
\be\label{kapa0}
\kappa_0=\lim_{\omega \rightarrow 0}\frac{I_{p}(\omega)}{\left|\mu(\omega)\right|^2}\,.
\ee
Recall that $\gamma_0$ and $\kappa_0$ must be related according to (\ref{flucdis}). Furthermore, away from the low frequency limit, there is a more general relation between these coefficients,
\begin{equation}\label{generalFG}
2 \mathbf{Re}(\gamma[\omega]) = \frac{\beta}{m} \kappa{(\omega)}\,,
\end{equation}
which is known as the second fluctuation-dissipation theorem \cite{Kubo:f-d_thm}. In appendix \ref{subsub3}, we will check the validity of (\ref{generalFG}) in our holographic setup.

\section{Holographic computation\label{hbmsec}}
In this section we turn our attention to the holographic realization of Brownian motion in de Sitter space. We first revisit some basics on string dynamics from the bulk point of view. Later on, we explicitly compute the random force correlator and the admittance and we comment on the implications of our holographic results in terms of the proposed phenomenological Langevin equation (\ref{genLE}).

\subsection{Heavy quarks and string dynamics\label{subsub1}}

In the context of AdS/CFT, the introduction of an open string sector
associated with a stack of $N_f$  flavor branes in the geometry (\ref{metric1}) is equivalent to the
addition of $N_f$ hypermultiplets in the fundamental representation of the gauge group, and these are the degrees
of freedom that we will refer to as quarks. For $N_f\ll N$, the backreaction on the geometry can be neglected
and, from the gauge theory perspective, this corresponds to working in a ``quenched approximation'' which disregards quark loops.

More specifically, a heavy quark on the boundary theory corresponds to the endpoint of an open string that stretches between the boundary, $z=0$, and the bulk horizon located at $z_h=2/H$. The dynamics of this string follows as usual from the Nambu-Goto action:\footnote{For a review of quark dynamics in the context of AdS/CFT see \cite{Chernicoff:2011xv}.}
\begin{equation}\label{NGAction}
S_{\text{NG}} = -\frac{1}{2\pi\alpha'} \int_\Sigma d\sigma d\tau \sqrt{-\mathrm{det} g_{\alpha\beta}}=\frac{1}{2\pi\alpha'} \int_\Sigma d\sigma d\tau\,\mathcal{L}_{\text{NG}}\,,
\end{equation}
where $g_{\alpha\beta}=G_{mn}\p_\alpha X^m\p_\beta X^n$ is the induced metric on the worldsheet $\Sigma$, and $X^{m}{(\tau,\sigma)}$ are the embedding functions of the string in the bulk spacetime. We will work in the static gauge $(\tau,\sigma)=(t,z)$.

One can easily verify that the embedding $X^m=(t,0,...,0,z)$ is a trivial solution and this correspond to a static quark that is placed at $r=0$. To obtain the rest-mass of the Brownian particle, we integrate the energy density of this static string. The canonical momentum densities are defined through
\be
\Pi_\mu^{\alpha}\equiv-\frac{1}{2\pi\alpha'}\frac{\p \mathcal{L}_{\text{NG}}}{\p(\p_\alpha X^{\mu})}\,,
\ee
and, in the static gauge, the energy density $\mathcal{E}=\Pi_t^{t}$ reads
\begin{equation}
\mathcal{E}=\frac{L^2}{2\pi\alpha'}\frac{f(z)}{z^2}\,,
\end{equation}
where
\be\label{fdez}
f(z)=1-\frac{H^{2}z^{2}}{4}\,.
\ee
Note that if the string extends all the way to the boundary at $z=0$, it would mean that the external particle is infinitely heavy and therefore it would not undergo Brownian motion. Thus, we impose a radial cutoff $z_m$ to render the mass finite.\footnote{This cutoff is fixed by the location of the flavor branes, which introduces finite mass (and hence dynamical) quarks into the boundary theory.}
Integrating along $z$ we obtain,
\begin{equation}
m = \frac{L^2}{2\pi\alpha'} \int_{z_m}^{z_h} \frac{dz}{z^2}\left(1-\frac{H^{2}z^{2}}{4}\right)= -\frac{L^2}{2\pi\alpha'}\left[\frac{1}{z}+\frac{H^{2}z}{4}\right]_{z_m}^{z_h} = \frac{\sqrt{\lambda}}{2\pi z_m}\left(1-\frac{Hz_m}{2}\right)^2\, ,
\end{equation}
where we have identified the 't Hooft coupling as $\lambda=L^4/\alpha'^{\,2}$.\footnote{This identification is correct only in the $d=4$ case. However, we will also use the same definition for other number of dimensions.} Notice that this integral blows up as $z_m\to0$, as anticipated. The $H$-dependent terms can be interpreted as the thermal correction to the mass. In terms of the temperature, $T_{\text{dS}}$, we have
\be
m =\frac{\sqrt{\lambda}}{2\pi z_m}\left(1-\pi T_{\text{dS}} z_m\right)^2\, ,
\ee
which can be inverted to obtain
\be
z_m=\frac{\sqrt{\lambda}}{2\pi m}\left(1-\frac{\sqrt{\lambda}T_{\text{dS}}}{m}+\mathcal{O}\left(\frac{\lambda T_{\text{dS}}^2}{m^2}\right)\right)\,.
\ee
The introduction of finite mass has important consequences from the field theory perspective.
In this case, the quark described by the string is not `bare' but `composite' or `dressed',
with a gluonic cloud of size $z_m$ \cite{Hovdebo:2005hm,Chernicoff:2011vn,Agon:2014rda}.
This in turn leads to various phenomenological signatures like, for example, the modified dispersion relation found in \cite{Chernicoff:2008sa,Chernicoff:2009re,Chernicoff:2009ff}, a lower rate of energy loss at early times \cite{Guijosa:2011hf}, the photon peak predicted in \cite{CasalderreySolana:2008ne} and the Cherenkov emission of mesons analyzed in \cite{CasalderreySolana:2009ch,CasalderreySolana:2010xh}. Note however that we are only allowed to treat the string semiclassically as long as it is
sufficiently heavy. We are then restricted to work in the limit $z_m/z_h\ll1$ or, equivalently, $m\gg \sqrt{\lambda}H$. In the boundary theory, this in turn implies that the energy transferred in each collision is much smaller than the rest energy the Brownian particle and therefore, the motion will always be non-relativistic.

\begin{figure}[t!]
\begin{center}
  \includegraphics[width=10cm]{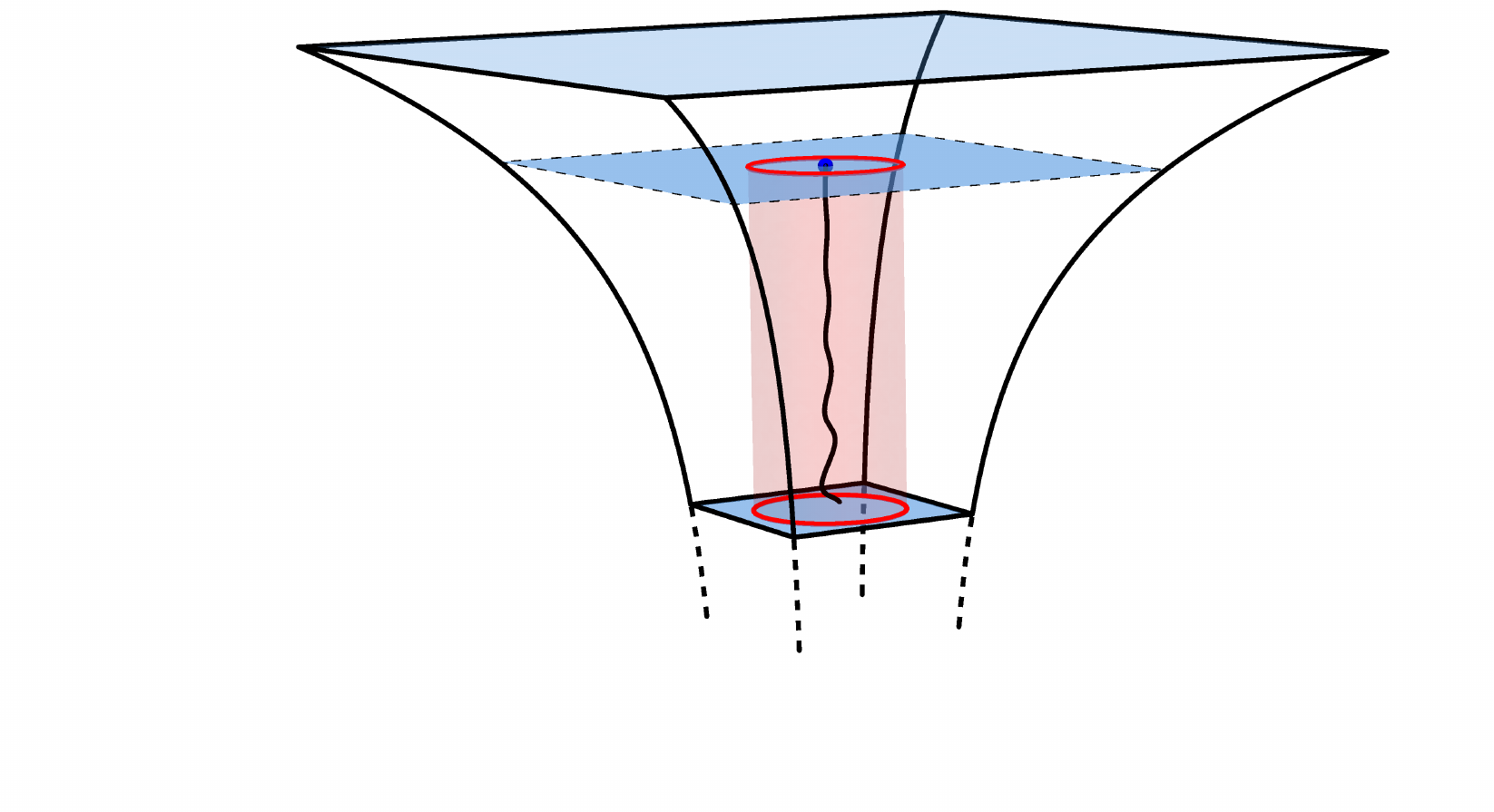}
\end{center}
\begin{picture}(0,0)
 \put(110,170){\small  $z=0$}
 \put(140,151){\small  $z=z_m$}
  \put(180,85){\small  $z=z_h$}
 \end{picture}
\vspace{-2.5cm}
\caption{Schematic picture of the
 string dual to a fluctuating heavy quark in de Sitter space.
The string stretches between the position of the flavor branes at $z = z_m$
and the bulk horizon at $z=z_h$. The cosmological horizon, depicted in red, is located at $r=1/H$ and extends into the bulk. \label{stringpic}}
\end{figure}
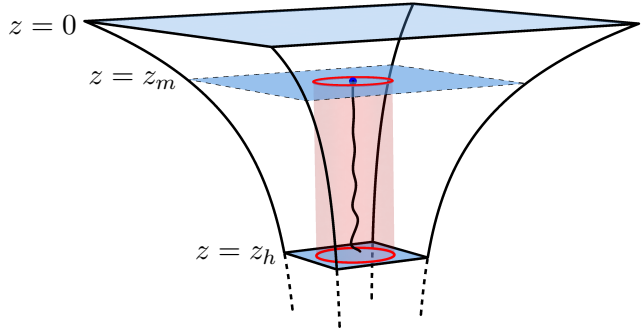

Let us now consider fluctuations around the static solution (see Figure \ref{stringpic} for a schematic picture
of the fluctuating string). For simplicity we will focus on fluctuations along one of the directions, say $x$, so the string embedding takes the form $X^{m}=(t,X(t,z),0,...,0,z)$. In this case the various components of the induced metric are found to be
\bea\label{wsmetric}
&&g_{tt} = \frac{L^{2}}{z^{2}}\left[-f(z)^2\left(1-H^{2}X^{2}\right) + \frac{f(z)^2}{1-H^{2}X^{2}}\dot{X}^{2}\right]\,,\nonumber\\
&&g_{zz} = \frac{L^{2}}{z^{2}}\left[1 + \frac{f(z)^2}{1-H^{2}X^{2}}X'^{2}\right]\,,\\
&&g_{tz} = \frac{L^{2}}{z^{2}}\frac{f(z)^2}{1-H^{2}X^{2}}\dot{X}X'\,,\nonumber
\eea
where $\dot{X}\equiv\p_t X$ and $X'\equiv \p_zX$. For general fluctuations, the action (\ref{NGAction}) contains higher order terms that would lead to nonlinear equations of motion. However, for small deviations from the origin, $XH\ll1$, and up to quadratic order in the perturbations\footnote{For fluctuations in $n\leq d-1$ spatial dimensions, $X^{m}=(t,X_1(t,z),...,X_{n}(t,z),0,...,0,z)$, the resulting action is just $n$ copies of (\ref{quadraticNG}) and the equations of motion for the various $X_i$ decouple. Therefore, without loss of generality we can focus in the $n=1$ case.}
\begin{equation}\label{quadraticNG}
S_{\text{NG}} \approx -\frac{\sqrt{\lambda}}{4\pi}\int dz dt\, \frac{f(z)}{z^{2}}\left[f(z)^2X'^{2}
-\dot{X}^{2}-H^{2}X^{2}\right]\,.
\end{equation}
Note that we have dropped the constant term that does not depend on $X$ and, therefore, will not contribute to the equations of motion.

We can also imagine the situation in which one has forced motion from the boundary perspective due to an external force. This can be easily realized in from the bulk perspective by turning on a world-volume $U(1)$ gauge field on the flavor branes. Since the
endpoint of the string is charged under this $U(1)$, this amounts to adding to the action the minimal coupling $S=S_{\text{NG}}+S_{\text{EM}}$, where
\be\label{actem}
S_{\text{EM}}=\int_{\p\Sigma}dt\left(A_t+A_x \dot{X}\right)\,.
\ee
This will exert the desired force on our Brownian particle. However, this is just a boundary term, so it will not play any role for the string dynamics in the bulk, other than modify the boundary condition. This will be important in section \ref{admittanceSEC} for the computation of the admittance. For now we will pay no attention to this part of the action.

The equation of motion coming from action (\ref{quadraticNG}) is
\begin{equation}\label{eom1}
\partial_{z}\left[\frac{f(z)^{3}}{z^{2}}X'\right] - \partial_{t}\left[\frac{f(z)}{z^{2}}\dot{X}\right] +\frac{f(z)}{z^{2}}H^{2}X = 0 \,.
\end{equation}
Now, given that the background (\ref{metric1}) is invariant under time translations, we can perform a Fourier decomposition
\be\label{fourier}
X(t,z)\sim e^{-i\omega t}g_\omega(z)\,.
\ee
Plugging (\ref{fourier}) in (\ref{eom1}) we obtain the following equation for the modes
\begin{equation}
\partial_{z}\left[\frac{1}{z^{2}}\left(1-\frac{H^{2}z^{2}}{4}\right)^3 g_\omega'(z)\right] + \frac{\omega^{2}+H^{2}}{z^{2}}
\left(1-\frac{H^{2} z^{2}}{4}\right)g_\omega(z) = 0\,.
\end{equation}
The two linearly independent solutions are
\bea
&& g^{(\text{in})}_{\omega}{(z)} = \frac{(4-4i\omega z + H^{2}z^{2})}{(4-H^{2}z^{2})}e^{i2\omega\, \mathrm{arctanh}(Hz/2)/H}\,,\label{insol}\\
&& g^{(\text{out})}_{\omega}{(z)} = g^{(in)}_{\omega}{(z)}^{*}\,,
\eea
where the superscripts $(\text{in})/(\text{out})$ refer to incoming and outgoing waves with respect to the bulk horizon. To see this notice that, around the unperturbed solution $X=0$, the worldsheet metric (\ref{wsmetric}) can be written as
\begin{equation}\label{wsmetric2}
ds^{2}_{\text{ws}} = \frac{L^2}{z^2}f(z)\left[- f(z)dt^{2}+\frac{dz^{2}}{f(z)}\right]\,,
\end{equation}
which is conformal to a metric of the ``black hole form''. This suggests that we can define a tortoise coordinate as
\begin{equation}\label{tortoisez}
z_{*} = - \int \frac{dz}{f(z)} = -\frac{2}{H}\mathrm{arctanh}{(Hz/2)}\,,
\end{equation}
where the horizon lies now at $z_{*}\to-\infty $.
In terms of this coordinate, the solutions above take the more familiar form
\bea
&& X^{(\text{in})}(t,z) \propto e^{-i\omega(t+z_{*})}\,,\\
&& X^{(\text{out})}(t,z) \propto e^{-i\omega(t-z_{*})}\,.
\eea
Note also that, in the near horizon limit, the action (\ref{quadraticNG}) can be written as
\begin{equation}
S_{\text{NG}} \propto\int dt dz_{*} \left[X'^{2}-\dot{X}^{2}-H^{2}X^{2}\right]\,,
\end{equation}
where now primes denote derivatives with respect to the tortoise coordinate $z_*$. This is precisely the form of the action
of a Klein-Gordon scalar field in Minkowski space with an inverted harmonic potential.

Away from the near horizon regime, however, the action (\ref{quadraticNG}) describes the dynamics of a scalar field $X(t,z)$ in a curved background (\ref{wsmetric2}). Standard quantization in a curved space \cite{bd,Wald:1995yp} leads to a mode expansion of the form
\begin{equation}
X(t,z) = \int_{0}^{\infty} \frac{d\omega}{2\pi}[a_{\omega}u_{\omega}(t,z)+a_{\omega}^{\dagger}u_{\omega}(t,z)^{*}]\,,
\end{equation}
where the functions $u_{\omega}$ correspond to a normalized basis of positive-frequency modes. These modes can be expressed as a linear combination of ingoing and outgoing waves with arbitrary coefficients,
\begin{equation}\label{modefnc}
u_{\omega}{(t,z)} = A_{\omega}[g^{(in)}_{\omega}{(z)}+B_{\omega} g^{(out)}_{\omega}{(z)}]e^{-i\omega t}\,.
\end{equation}
The constant $B_\omega$ is fixed through the boundary condition at $z=z_m$. In particular, in the absence of external forces one imposes a Neumann boundary condition
\begin{equation}
\partial_{z}u_{\omega}(t,z)|_{z = z_{m}} = 0\,,
\end{equation}
which leads to
\begin{equation}
B_\omega = -e^{4i\omega\,\mathrm{arctanh}(Hz_{m}/2)/H}\,.
\end{equation}
In particular, notice that this is a pure phase $\sim e^{i\theta}$. Then, outgoing and ingoing modes have the same amplitude and this implies that the black hole, which emits Hawking radiation, can be in thermal equilibrium \cite{Hemming:2000as}. On the other hand, the constant $A_\omega$ is obtained by requiring the normalization of the modes through the standard Klein-Gordon inner product. After some algebra (see Appendix \ref{normalizationSEC}) we get
\begin{equation}\label{constantA}
A_{\omega} = \left(\frac{\pi}{\sqrt{\lambda}\omega(H^{2}+\omega^{2})}\right)^{\!\!\frac{1}{2}}\,.
\end{equation}

The rest of this section is organized as follows. In \ref{subsub2} we compute the two-point functions of position and momentum, and from these we extract the displacement squared $\left\langle s(t)^2\right\rangle$ and the random force strength $\kappa(\omega)$, respectively. In \ref{admittanceSEC} we determine the admittance $\mu(\omega)$ and the friction kernel $\gamma(\omega)$ by measuring the response $\left\langle p(\omega)\right\rangle$ due to an external force. Finally, in \ref{remarks} we comment on the proposed Langevin equation (\ref{genLE}) and the time scales characterizing the evolution of the Brownian particle.

\subsection{Fluctuation: two-point functions\label{subsub2}}

In the semi-classical treatment, string modes are thermally excited by Hawking radiation emitted by the worldsheet horizon. In
particular, assuming that the state of the embedding fields is the usual Hartle-Hawking (or Euclidean) vacuum \cite{Lawrence:1993sg,Frolov:2000kx}, these modes follow the standard Bose-Einstein distribution:
\begin{equation}
\langle a_{\omega}^{\dagger} a_{\omega'} \rangle = \frac{2\pi\delta(\omega-\omega')}{e^{\beta\omega}-1}\,,\qquad \langle a_{\omega} a_{\omega'} \rangle = \langle a^{\dagger}_{\omega} a^{\dagger}_{\omega'} \rangle = 0\,,
\end{equation}
where
\begin{equation}
\beta = \frac{1}{T_{\text{dS}}} = \frac{2\pi}{H}\,.
\end{equation}
The position of the particle is identified as the string endpoint at $z=z_m$:
\begin{equation}
x(t) = X(t,z_{m}) = \int_{0}^{\infty} \frac{d\omega}{2\pi} [a_{\omega}u_{\omega}(t,z_{m})+a_{\omega}^{\dagger}u_{\omega}^{*}(t,z_{m})]\,.
\end{equation}
Substituting our solution for the modes into the above expression, we find
\begin{equation}\label{position}
x(t) = -2i z_{m} \int_{0}^{\infty} \frac{d\omega}{2\pi} \left[a_{\omega} \omega A_{\omega}  \left(\frac{2+Hz_{m}}{2-Hz_{m}}\right)^{i\frac{\omega}{H}} e^{-i\omega t} + \mathrm{h.c.}\right]\,,
\end{equation}
where $A_\omega$ is the normalization constant given in (\ref{constantA}). Next, we compute the 2-point function $\langle x(t)x(0) \rangle$. This has the usual IR divergence that comes from the zero-point energy, which exists even in the zero temperature limit $\beta\to \infty$. To avoid this, we regularize the correlator by implementing the normal ordering ${:\! a_\omega a_\omega^\dagger \!:} \equiv {:\! a_\omega^\dagger
a_\omega\! :}=a_\omega^\dagger a_\omega$. This yields\footnote{We have to keep in mind that normal-ordering does not respect the KMS conditions except in the classical limit. A different way to treat this divergence is by means of the canonical correlator--- see appendix \ref{subsub3} for details.}
\begin{equation}\label{2pointfunctionx}
\langle :\!x(t)x(0)\!: \rangle = \frac{\sqrt{\lambda}}{\pi^2m^2} \int_{0}^{\infty} d\omega \frac{\cos{(\omega t)}}{e^{\beta\omega}-1}\frac{\omega}{H^{2}+\omega^{2}}\,.
\end{equation}
From this correlator, we can compute the displacement squared. After some straightforward algebra, we obtain
\begin{equation}\label{s2mid}
\langle :\!s^{2}(t)\!:\rangle = \frac{4\sqrt{\lambda}}{\pi^2m^2} \int_{0}^{\infty} d\omega \frac{\sin^{2}{(\omega t/2)}}{e^{\beta\omega}-1}\frac{\omega}{H^{2}+\omega^{2}}\,,
\end{equation}
where in the process we used the identity $1-\cos{(x)}=2\sin^{2}{(x/2)}$. Expression (\ref{s2mid}) can be written in terms of a dimensionless integral as follows:
\begin{equation}\label{s2}
\langle :\!s^{2}{(t)}\!:\rangle = \frac{\sqrt{\lambda}}{\pi^{4}m^{2}} \int_{0}^{\infty} dx \frac{\sin^{2}{(\frac{Ht}{4\pi}x)}}{e^{x}-1}\frac{x}{1+\frac{x^2}{4\pi^{2}}}\,.
\end{equation}
Notice that the displacement squared vanishes in two limits: $m \rightarrow \infty$ and $H \rightarrow 0$. In the first case, the particle is infinitely heavy and, therefore, we do not expect it to undergo Brownian motion; in the second case, the temperature vanishes and the particle is not expected to undergo Brownian motion either.

We explicitly evaluate the integral in (\ref{s2}) in appendix \ref{s2app}. The final result is
\bea\label{finals2}
&&\langle :\!s^{2}{(t)}\!:\rangle = \frac{\sqrt{\lambda}}{2\pi^{2}m^{2}}\big[2\gamma_{E}-e^{Ht}\mathrm{Ei}{(-Ht)}-e^{-Ht}\mathrm{Ei}{(Ht)} \nonumber\\
&&\qquad\qquad\qquad\qquad\qquad+ (e^{Ht}+e^{-Ht})\log{(1-e^{-Ht})}+Ht e^{-Ht}\big]\,,
\eea
where $\gamma_{E}$ is the Euler-Mascheroni constant and ${\rm Ei}(z)$ is the exponential integral. At early times, we recover the anticipated ballistic regime,
\begin{equation}\label{ballistic}
\langle :\!s^{2}{(t)}\!:\rangle \approx v_{0}^{2}t^{2} + \mathcal{O}((Ht)^{4})\,,
\end{equation}
where the constant $v_{0}$ is given by
\begin{equation}\label{v0s2}
v_{0}^{2} = \left(\frac{\sqrt{\lambda}H^{2}}{4\pi^{2}m^{2}}\right)\left(\frac{7}{6}-2\gamma_{E}\right)\,.
\end{equation}
However, in the scaling of (\ref{v0s2}), there is an extra factor of order $\sim \mathcal{O}(\sqrt{\lambda}T_{dS}/m)$ in comparison to the expected behavior in thermal field theories. This peculiar behavior implies that the energy transferred in a single collision is $ \sim \sqrt{\lambda}T_{\text{dS}}^2/m$, instead of instead of the standard $\sim T_{\text{dS}}$ predicted by the equipartition theorem. We will comment more on this point in appendix \ref{subsub4}, where we explicitly compute the distribution of velocities of the Brownian particle.

In terms of $v_{0}$, the displacement squared takes the form
\begin{eqnarray}
&&\langle :\!s^{2}{(t)}\!:\rangle = \frac{v_{0}^{2}}{H^{2}\left(\frac{7}{12}-\gamma_{E}\right)} \big[2\gamma_{E}-e^{Ht}\mathrm{Ei}{(-Ht)}-e^{-Ht}\mathrm{Ei}{(Ht)} \nonumber \\
&&\qquad\qquad\qquad\qquad\qquad\qquad\quad +(e^{Ht}+e^{-Ht})\log{(1-e^{-Ht})}+Ht e^{-Ht}\big]\,.
\end{eqnarray}
In the late time regime, we find that the displacement squared does, indeed, approach a constant value:
\begin{equation}\label{latetimes2}
\langle :\!s^{2}{(t)}\!:\rangle = \frac{v_{0}^{2}}{H^2}\frac{(2\gamma_{E}-1)}{(\frac{7}{12}-\gamma_{E})} + \mathcal{O}((Ht)^{-2})\,.
\end{equation}
Since the Brownian particle is very massive, its motion is non-relativistic ($v_0 \ll 1$) and, in the limit above, $\langle :\!s^{2}{(t)}\!:\rangle$ is in general smaller than $1/H^2$. If we take $v_{0}$ to be at most one order of magnitude smaller than the speed of light then, according to (\ref{latetimes2}), the furthest the particle can go is approximately halfway to the horizon. Notice that this is what the observer at $r=0$ sees, as the coordinate $t$ measures the time according to such an observer. Intuitively, the fact that the particle does not actually reach the horizon is due to the fact that it interacts with more and more Hawking quanta as it approaches the horizon. At the equilibrium position then, the random kicks counteract the gravitational repulsion due to the de Sitter metric.

\begin{figure}[t!]
$$
\begin{array}{ccc}
  \includegraphics[width=4.8cm]{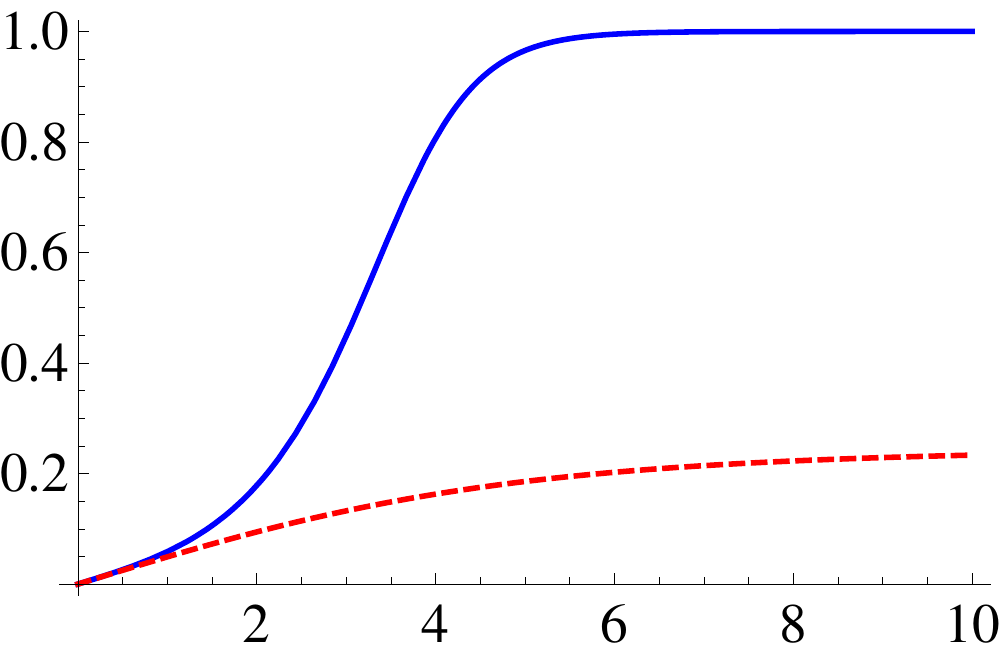} & \;\; \includegraphics[width=4.8cm]{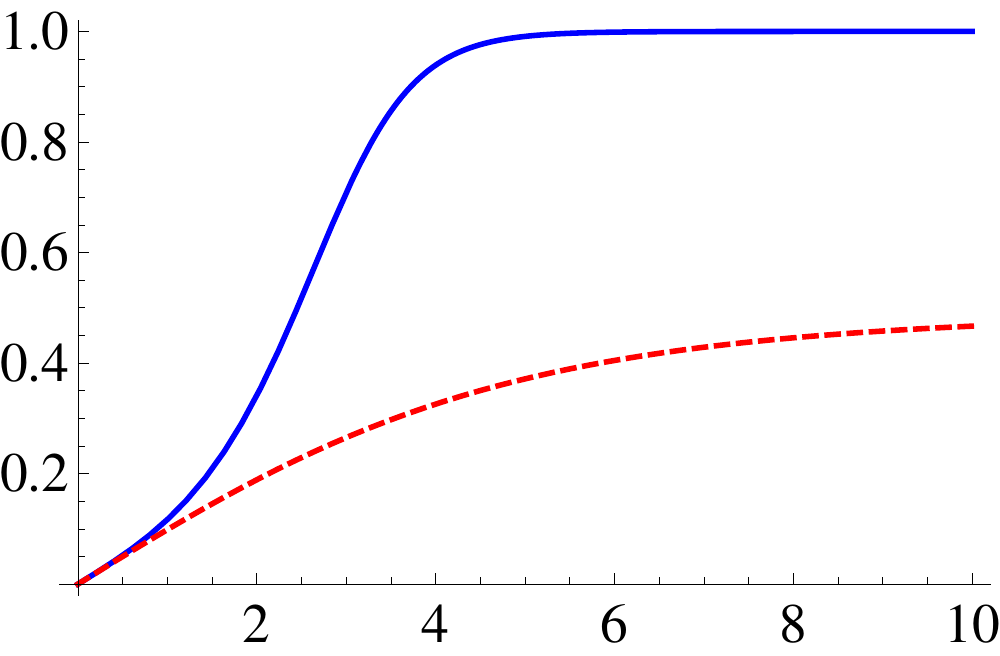} & \;\; \includegraphics[width=4.8cm]{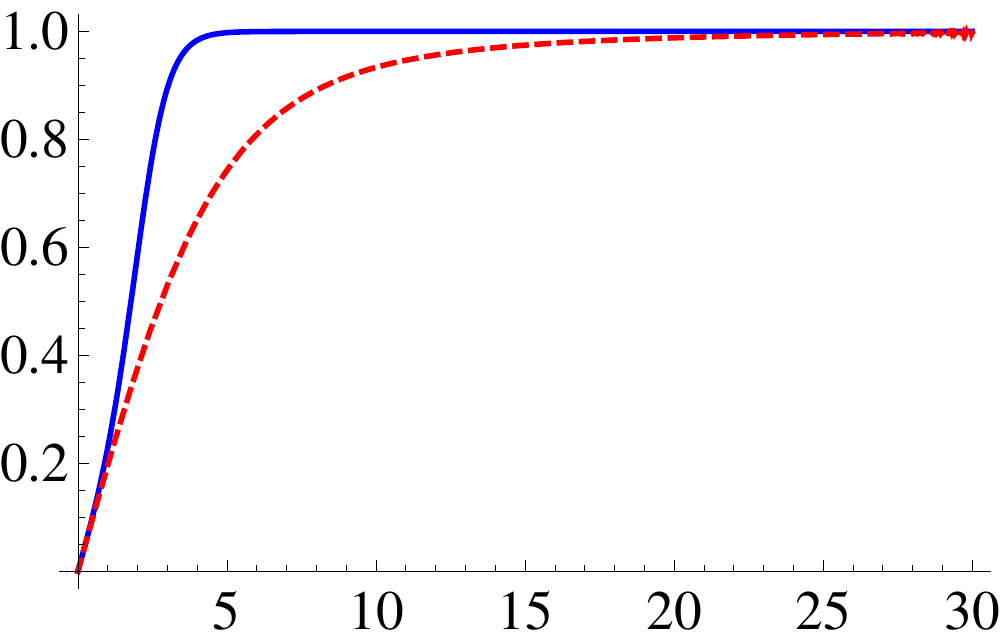}
\end{array}
$$
\begin{picture}(0,0)
 \put(2,110){\footnotesize  $r(t),s(t)$}
 \put(146,25){\footnotesize  $t$}
  \put(149,110){\footnotesize  $r(t),s(t)$}
 \put(293,25){\footnotesize  $t$}
  \put(296,110){\footnotesize  $r(t),s(t)$}
 \put(441,25){\footnotesize  $t$}
 \end{picture}
\caption{Comparison of $r(t)$ predicted by the geodesic equation (blue solid curve) vs. $s(t)\equiv\sqrt{\langle s^2(t)\rangle}$ (red dashed curve) for $v_0=5\times10^{-2}$ (left), $v_0=10^{-1}$ (center) and $v_0=2\times10^{-1}$ (right). Here we have set $H=1$. For $v_0\gtrsim5\times10^{-2}$ the approximation breaks down at late times, given that $s(t)$ becomes of order $\mathcal{O}(1)$
\label{plotsrs}}
\end{figure}
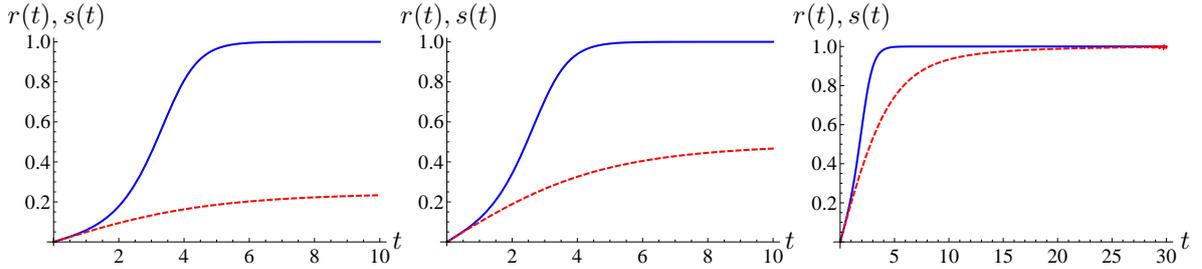

From the plots of the radial geodesic and the displacement squared (see Figure \ref{plotsrs}) we can see that, in both cases, the particle remains inside the horizon even after infinitely long time. It is important to keep in mind, however, that $t$ is only a coordinate time. In fact, from the results of appendix \ref{geoapp} we can see that the radial geodesic does exit the horizon after a finite \emph{proper} time. This is best understood by a close inspection of Figure \ref{static}. From this figure, it is apparent that there are curves with finite proper lengths from the center to the horizon. Any curve reaching the horizon, however, must necessarily intersect infinitely many black curves (\emph{i.e.} curves of constant $t$). This is why the radial geodesic does exit the horizon even though it takes infinitely long coordinate time to reach it. On the other hand, it is quite clear that the Brownian particle stays well inside the static patch.

Let us now focus on the momentum correlator,
\begin{equation}\label{ppcor}
\langle :\!p(t)p(0)\!: \rangle = -m^{2} \partial_{t}^{2} \langle :\!x(t)x(0)\!: \rangle = \frac{\sqrt{\lambda}}{\pi^2} \int_{0}^{\infty} d\omega \frac{\cos{(\omega t)}}{e^{\beta\omega}-1}\frac{\omega^{3}}{H^{2}+\omega^{2}}\,.
\end{equation}
This can be rewritten as
\begin{equation}
\langle :\!p(t)p(0)\!: \rangle = \frac{\sqrt{\lambda}}{2\pi^2} \int_{-\infty}^{\infty} d\omega \frac{e^{-i\omega t}}{e^{\beta|\omega|}-1}\frac{|\omega|^{3}}{H^{2}+\omega^{2}}\,,
\end{equation}
from which we can read the power spectrum,
\begin{equation}\label{Ip}
I_{p}{(\omega)} =\frac{\sqrt{\lambda}}{\pi} \left(\frac{1}{e^{\beta|\omega|}-1}\right)\left(\frac{|\omega|^{3}}{H^{2}+\omega^{2}}\right)\,.
\end{equation}
Finally, using (\ref{kapa}) we can also extract the random force strength,
\begin{equation}\label{kappaomega}
\kappa{(\omega)} = \frac{\sqrt{\lambda}}{\pi} (H^{2}+\omega^{2})\frac{|\omega|}{e^{\beta|\omega|}-1}\,,
\end{equation}
which, in the low frequency limit, yields
\begin{equation}\label{kappa0}
\kappa_{0} = \frac{\sqrt{\lambda}H^{3}}{2\pi^{2}}=4\pi\sqrt{\lambda}T^3_{\text{dS}}\,.
\end{equation}
Note that (\ref{kappa0}) is the expected result in a thermal field theory in $1+1$ dimensions. More in general, the holographic result for a Schwarzschild-AdS$_{d+1}$ background is \cite{brownian}
\be
\kappa_{0}^{d+1}=\frac{16\pi\sqrt{\lambda}T^3}{d^2}\,.
\ee
The fact that, for de Sitter space, $\kappa_{0}$ turns out to be independent of the number of dimensions is a surprising result of the AdS/CFT correspondence. This result is not obvious at all. On one hand, the thermodynamics of the hyperbolic black holes dual to the theories considered here do depend on the number of dimensions \cite{Emparan:1998he,Birmingham:1998nr,Emparan:1999gf}. In addition, the holographic stress-energy tensor is also found to depend on $d$ as shown in \cite{Marolf:2010tg,Fischler:2013fba}. Nevertheless, this particular behaviour could be anticipated from the form of the metric (\ref{metric1}) itself. In particular, note that the function $f(z)$ is independent of the number of dimensions, whereas in a Schwarzschild-AdS black hole the power of the $z$-term depends explicitly on $d$.\footnote{Other thermal properties of these theories are also found to be independent of the number of dimensions \cite{progress}.}

\subsection{Dissipation: the response function\label{admittanceSEC}}
Let us now study the response of the system due to an external force $F(t) \sim  e^{-i \omega t} F(\omega)$.
In the linear response theory, this is characterized by the retarded Green's function $\chi(\omega)$ through
\begin{equation}
\langle x(\omega) \rangle = \chi(\omega) F(\omega)\,.   \label{response}
\end{equation}
In the gravity side, the external force is introduced by turning on an electric field on the flavor branes, $F_{xt} =F(t)$. This amounts the addition of the boundary term (\ref{actem}) to the string action, which in terms of $F(t)$ becomes
\begin{equation}
S_{\text{EM}} = \int dt dz\,\delta(z-z_m) F(t) X(t,z)\,.
\end{equation}
This modifies the UV boundary condition according to
\begin{equation}\label{UVBC}
-\frac{1}{2\pi\alpha'}\left. \frac{\partial \mathcal{L}_{\text{NG}}}{\partial X'} \right |_{z_m} =  \frac{\sqrt{\lambda}}{2\pi} \frac{f(z)^{3}}{z^2} X'(t,z)\bigg|_{z_m} = F(t)\, .
\end{equation}
On the other hand, the retarded Green's function $\chi(\omega)$ selects an ingoing boundary condition in the IR \cite{Son:2002sd,Herzog:2002pc}. This can be explained as follows:
the general solution for $X$ is the sum of ingoing and outgoing waves at the horizon $X= A^{(\text{in})}X^{(\text{in})}+A^{(\text{out})} X^{(\text{out})}$, where $X^{(\text{in/out})}=e^{-i\omega t}g_\omega^{(\text{in/out})}$. In the semiclassical treatment, the outgoing modes are thermally excited by Hawking radiation, while the ingoing
modes can be arbitrary. In addition, given that the Hawking radiation is random, the phase of $A^{(\text{out})}$ is randomly distributed
and, taking the statistical average we get $\langle A^{(\text{out})}\rangle=0$.\footnote{On the other hand, the advanced Green's function selects the outgoing boundary condition at the horizon.} We can thus write
\be
\langle x(t)\rangle\equiv\langle X(t,z)\rangle\big|_{z_m} = \langle A^{(\text{in})}\rangle e^{-i\omega t}g_\omega^{(\text{in})}(z)\big|_{z_m}\,,
\ee
where $g^{(\text{in})}(z)$ corresponds to the ingoing solution given in (\ref{insol}).

Now, from (\ref{UVBC}) it follows that
\begin{equation}
F(\omega) =  \langle A^{(\text{in})}\rangle \frac{\sqrt{\lambda}}{2\pi}  \frac{f(z)^{3}}{z^2}  g'^{(\text{in})}_\omega(z)\bigg|_{z_m}\,.
\end{equation}
and from (\ref{response}) we get
\begin{equation}\label{chi}
\chi{(\omega)} =\frac{2\pi}{\sqrt{\lambda}} \frac{z^2 g_\omega^{(\text{in})}(z)}{f(z)^{3}g'^{(\text{in})}_\omega(z)}\bigg|_{z_m}= -\frac{2\pi}{\sqrt{\lambda}}\left(\frac{z_{m}-i\omega z_{m}^{2}}{H^{2}+\omega^{2}}\right) + ... \,,
\end{equation}
where the dots denote terms that are higher order in $z_m$. Finally, the admittance $\mu(\omega)$, defined in (\ref{admitt}), is related to the response function through
\begin{equation}\label{mu}
\mu{(\omega)} = -im\omega \chi{(\omega)} =\frac{2\pi m}{\sqrt{\lambda}}\left(\frac{i\omega z_{m} + \omega^{2}z_{m}^{2}}{H^{2}+\omega^{2}}\right)\,.
\end{equation}

In the low frequency limit, this result implies through (\ref{frictioncoeff}) that
\begin{equation}\label{gamma0}
\gamma_{0} = \frac{\sqrt{\lambda}H^{2}}{2\pi m}=\frac{2\pi\sqrt{\lambda}T^2_{\text{dS}}}{m}\,,
\end{equation}
and
\begin{equation}\label{koverm}
\frac{k}{m} = -H^{2}\,.
\end{equation}
A few comments are in order here. First note that, similar to the result for the random force correlator, (\ref{gamma0}) is the expected result for thermal field theories in $1+1$ dimensions, which in general yields \cite{brownian}
\be
\gamma_{0}^{d+1}=\frac{8\pi\sqrt{\lambda}T^2}{d^2 m}\,.
\ee
On the other hand, note that (\ref{koverm}) is precisely the expected behavior for de Sitter space. Therefore, this result provides a non-trivial check that the AdS/CFT calculation is indeed consistent with the Brownian motion of a particle in the inverted harmonic potential (\ref{pot}).
Finally, note also that from (\ref{gamma0}) and (\ref{kappa0}), we can see that the fluctuation-dissipation theorem holds, at least in its weak form (\ref{flucdis}). More in general, the fluctuation-dissipation theorem is expected to be valid for all frequencies, according to (\ref{generalFG}). However, as mentioned in the previous section, normal-ordering only preserves the KMS relations in the classical limit and hence the results for $\kappa(\omega)$ will get corrections for finite $\omega$. In appendix \ref{subsub3} we will choose a different regularization scheme and we will explicitly show that the second fluctuation-dissipation relation (\ref{generalFG}) is indeed valid in our holographic setup.

Before proceeding further, let us write down the result for $\gamma{(\omega)}$. From (\ref{frictioncoefff}) we get
\begin{equation}\label{gammaomega}
\gamma{(\omega)} = \frac{(H^{2}+\omega^{2})(1+i\omega z_{m})z_{m}}{1+\omega^{2}z_{m}^{2}}\,.
\end{equation}
We will analyze the physical implications of (\ref{gammaomega}) and (\ref{kappaomega}) in the next subsection.

\subsection{Time scales and the holographic Langevin equation\label{remarks}}
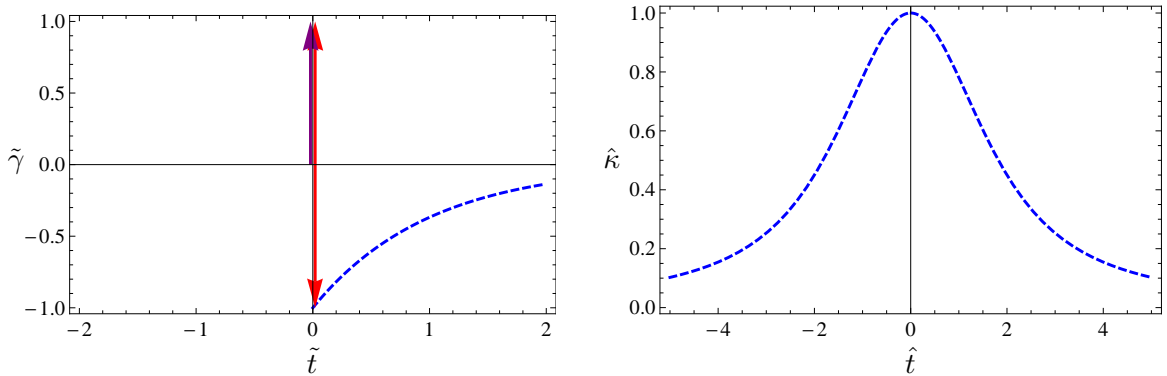
\begin{figure}[t!]
$$
\begin{array}{cc}
  \includegraphics[width=7cm]{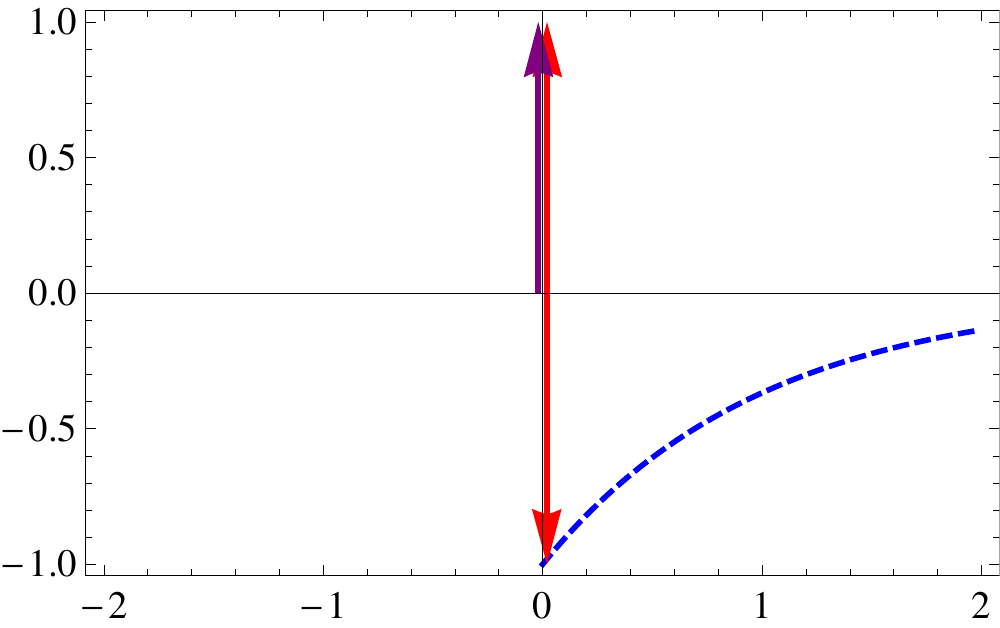} & \quad\quad \includegraphics[width=7cm]{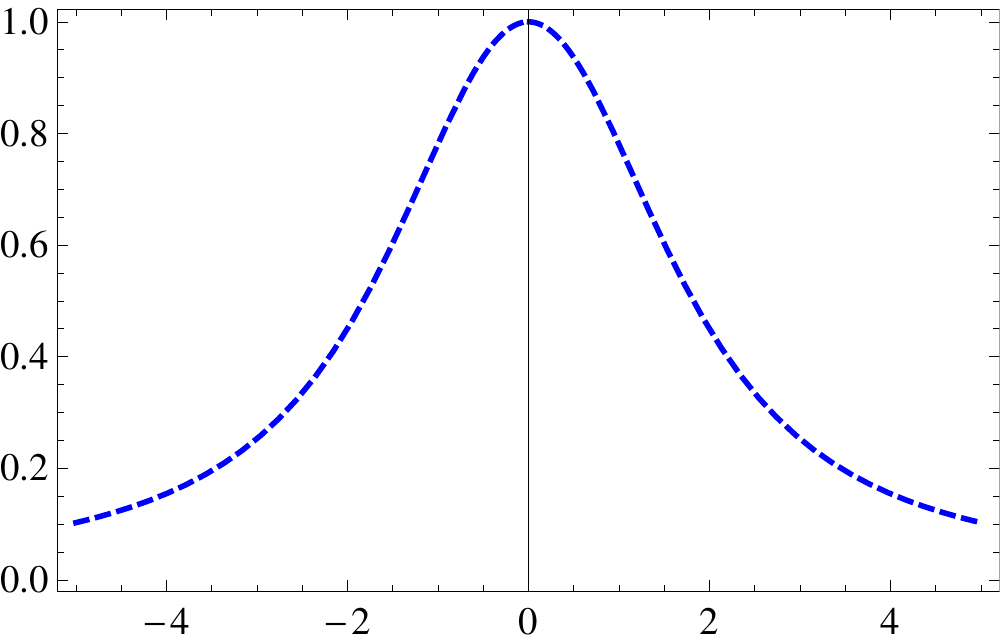}
\end{array}
$$
\begin{picture}(0,0)
 \put(5,83){\small $\tilde{\gamma}$}
 \put(117,6){\small $\tilde{t}$}
  \put(228,83){\small $\hat{\kappa}$}
 \put(341,6){\small $\hat{t}$}
 \end{picture}
\caption{Left panel: dimensionless $\tilde{\gamma}$ as a function of $\tilde{t}$. The purple arrow in this plot denotes the delta function. The double red arrow, on the other hand, represents the derivative of the delta function. Right panel: dimensionless $\hat{\kappa}$ as a function of $\hat{t}$.\label{plots}}
\end{figure}

Given the results of sections \ref{subsub2} and \ref{admittanceSEC}, the aim of this section is to comment on the implications in terms of the generalized Langevin equation proposed in (\ref{genLE}). First, let us start by Fourier transforming the friction kernel (\ref{gammaomega}) and the random force strength (\ref{kappaomega}), to obtain the real-time coefficients
\begin{equation}\label{coefficients}
\gamma{(t)} = \int_{-\infty}^{\infty} \frac{d\omega}{2\pi} \gamma{(\omega)} e^{-i\omega t}\quad\text{and}\quad\kappa{(t)} = \int_{-\infty}^{\infty} \frac{d\omega}{2\pi} \kappa{(\omega)} e^{-i\omega t}\,.
\end{equation}
The first one is straightforward, and yields
\begin{equation}\label{gammat}
\gamma{(t)} = \frac{2\pi m}{\sqrt{\lambda}}\delta{(t)} - \delta'{(t)} - \frac{4\pi^2m^2}{\lambda}\left(1-\frac{\lambda H^2}{4\pi^2m^2}\right)e^{-2\pi m t/\sqrt{\lambda}}\theta{(t)}\,.
\end{equation}
Here $\theta(t)$ is the unit step function, $\delta(t)$ is the Dirac delta function and $\delta'(t)$ its derivative. As we can see, for $t < 0$ we have $\gamma(t) = 0$, as required by causality.
The second coefficient in (\ref{coefficients}) requires a little more work. We first rewrite it as
\begin{equation}\label{kappat}
\kappa{(t)} = \sqrt{\lambda}H^{4}\left[\frac{1}{8\pi^{3}}K_{1}+\frac{1}{32\pi^{5}}K_{2}\right]\,,
\end{equation}
where
\begin{equation}
K_{1} = \int_{-\infty}^{\infty} dx \frac{|x|}{e^{|x|}-1}e^{-iHtx/2\pi}\quad\text{and}\quad K_{2} = \int_{-\infty}^{\infty} dx \frac{|x|^3}{e^{|x|}-1}e^{-iHtx/2\pi}\,.
\end{equation}
The integral $K_{1}$ was computed in \cite{brownian} and evaluates to
\begin{equation}
K_{1} = \left(\frac{2\pi}{H t}\right)^{2} - \frac{\pi^{2}}{\sinh^{2}{(H t/2)}}\,.
\end{equation}
Also, it is easy to see that
\begin{equation}
K_{2} = -\beta^{2}\frac{\partial^{2}K_{1}}{\partial t^{2}}=\frac{2\pi^{4}}{\sinh^{4}{(H t/2)}} + \frac{4\pi^{4}}{\tanh^{2}{(H t/2)}\sinh^{2}{(H t/2)}} - 6\left(\frac{2\pi}{Ht}\right)^4\,.
\end{equation}
We can now rewrite (\ref{gammat}) and (\ref{kappat}) in dimensionless form. In the limit $\sqrt{\lambda}H/m\ll1$ we can ignore the second term proportional to $\theta(t)$ and then, defining $\tilde{\gamma} = 4\pi^2m^2\gamma/\lambda$ and $\tilde{t} = 2\pi mt/\sqrt{\lambda}$,
\begin{equation}\label{tildeg}
\tilde{\gamma}{(\tilde{t})} = \delta{(\tilde{t})} - \delta'{(\tilde{t})} -e^{-\tilde{t}}\theta{(\tilde{t})}\,.
\end{equation}
For the random force strength we choose $\hat{\kappa} = \kappa(t)/\kappa(0)=240 \pi \kappa/11\sqrt{\lambda}H^{4}$ and $\hat{t}=Ht/2$. This gives
\begin{equation}\label{hatk}
\hat{\kappa}(\hat{t}) = \frac{30}{11}\left(\frac{1}{\hat{t}^{2}}-\frac{1}{\sinh^{2}{\hat{t}}}\right) + \frac{15}{11}\left(\frac{1}{\sinh^{4}{\hat{t}}}+\frac{2}{\tanh^{2}{\hat{t}}\sinh^{2}{\hat{t}}}-\frac{3}{\hat{t}^{4}}\right)\,.
\end{equation}
In Figure \ref{plots} we can see plots of (\ref{tildeg}) and (\ref{hatk}) respectively.

The Langevin equation in time domain is then:
\begin{equation}
2\dot{p}(t) = -\frac{2\pi m }{\sqrt{\lambda}}p(t) + \frac{4\pi^2m^2}{\lambda}\left(1-\frac{\lambda H^2}{4\pi^2m^2}\right) \int_{-\infty}^{t} dt' e^{-2\pi m(t-t')/\sqrt{\lambda}}p{(t')} +mH^2x(t)+ R(t) \,,
\end{equation}
where we have integrated the $\delta'(t)$ term by parts. Note that the noise kernel has an exponential tail, a very commonly used kernel in the literature. It is interesting that extra terms that are independent of $H$ appear in the above equation. These come from the delta term in (\ref{gammat}), the derivative of the delta and the $H$-independent piece that appears in front of the $\theta(t)$ function. This is counterintuitive at first sight. However, we believe these extra terms are precisely the responsible for the anomalous behavior of the Brownian particle at late times.

We can also compute the time scales that characterize the evolution of the Brownian particle. From the low frequency limit of the friction coefficient we can read the relaxation as time defined in (\ref{t_relax}). In our case we obtain
\be\label{trel}
t_{\text{relax}}\sim\gamma_0^{-1}\sim\frac{m}{\sqrt{\lambda}T^2_{\text{dS}}}\,.
\ee
This essentially sets the time scale for which the Brownian particle loses memory of its initial conditions and thermalizes, \emph{i.e.} the time at which the particle enters the regime of slow anomalous diffusion. Another relevant time scale is the collision time $t_{\text{coll}}$, which is taken to be the width of the random force function $\kappa(t)$. More specifically, we can define the collision time as
\begin{equation}\label{tcolldef}
t_{\text{coll}} = \int_{0}^{\infty} dt \frac{\kappa{(t)}}{\kappa{(0)}}\,,
\end{equation}
so that, if $\kappa(t) = \kappa(0)e^{-t/t_{\text{coll}}}$, equation (\ref{tcolldef}) gives precisely $t_{\text{coll}}$. After a quick calculation we obtain
\begin{equation}\label{tcoll}
t_{\text{coll}} = \frac{30}{11\pi T_{\text{dS}}} \sim \frac{1}{T_{\text{dS}}}\,,
\end{equation}
which is qualitatively what we expected from Figure \ref{plots}. This quantity measures the time scale over which the random force $R(t)$ is correlated, and can be interpreted as the
duration of a single scattering event. In typical examples of Brownian motion
\begin{align}
 t_{\text{relax}}\gg t_{\text{coll}}\,.
\label{tr>>tc}
\end{align}
This includes the case where the fluid is very dilute so the Brownian particle is not scattered very often, and the case where the fluid constituents are much smaller than the Brownian particle itself \cite{Kubo:f-d_thm}. In our case, this statement holds true in the regime of validity of our computations, \emph{i.e.} for $m/\sqrt{\lambda}T_{\text{dS}}\gg1$.
Note that both, (\ref{trel}) and (\ref{tcoll}) agree with the results of \cite{brownian,Atmaja:2010uu} for finite-temperature field theories with a gravity dual.
Additionally, there is also a third scale $t_{\text{mfp}}$ that is interpreted as the typical time lapse between two consecutive
collisions. However, the computation of this quantity requires the evaluation of a four-point function \cite{brownian,Atmaja:2010uu}, which is beyond the scope of this paper. Nevertheless, we expect $t_{\text{mfp}}$ to be at least of order\footnote{The computation of $t_{\text{mfp}}$ in \cite{Atmaja:2010uu} gives a somehow larger result at strong coupling: $t_{\text{mfp}}\sim1/T\log\lambda\,.$}
\be\label{tmfp}
t_{\text{mfp}}\sim\frac{1}{\sqrt{\lambda}T_{\text{dS}}}\,.
\ee
This follows from a simple physical argument: if we set $ m = T_{\text{dS}}$ then $t_{\text{relax}}$ should give $t_{\text{mfp}}$, because any given
particle of the thermal bath can be considered as a Brownian particle of mass $\sim T_{\text{dS}}$. In typical examples of Brownian motion, the mean free path time $t_{\text{mfp}}$ is such that $t_{\text{coll}}\ll t_{\text{mfp}} \ll t_{\text{relax}}$. However, if (\ref{tmfp}) holds true, this separation is no longer valid.
The fact that $t_{\text{coll}}\gg t_{\text{mfp}}$ implies that, in the limit $\lambda\to\infty$, a Brownian particle interacts with many fluid constituents
at the same time. As a result, a particle of mass $m \sim \sqrt{\lambda} T_{\text{dS}}$ can thermalize much faster than usual,
$t_{\text{relax}}\sim t_{\text{coll}}$, \emph{i.e.} after a single collision.

\section{Conclusions and outlook\label{sec5}}

In this paper we have studied various thermal properties of QFTs in the static patch of de Sitter space, assuming that the quantum fields are in the Bunch-Davies (or Euclidean) vacuum. The models in consideration have gravity duals coming from the standard Einstein-Hilbert action with negative cosmological constant and, hence, can be thought of as theories that belong to a universality class of strongly-coupled CFTs in the large-$N$ limit. More specifically, we considered a heavy quark that interacts with Hawking radiation (emitted by the cosmological horizon) and undergoes an erratic motion.  From the point of view of AdS/CFT, the heavy quark is realized by the introduction of an open string that stretches between the boundary and the bulk horizon. The modes of the string are then thermally excited by the Hawking radiation (emitted by the bulk horizon) and, as a consequence, the endpoint of the string dual to the heavy quark fluctuates randomly. This behaviour exhibits many features reminiscent of the Brownian motion of a heavy particle in an inhomogeneous medium.

One of the more interesting results of this paper is the behavior of the displacement squared $\langle s^2(t)\rangle$, both in the early time and late time regimes (\ref{finals2}). We found that, similar to ordinary Brownian motion, the heavy quark exhibits a ballistic regime at early times (\ref{ballistic}), but its velocity squared is different from that predicted by equipartition of energy (\ref{v0s2}). This issue was studied in more detail in appendix \ref{subsub4} where we explicitly evaluated the distribution of momentum. As a result, we showed that $f(p)$ follows a Gaussian distribution, as expected by time translation invariance, but with a width that differs from the standard Maxwell-Boltzmann distribution (\ref{disfp}). This implies that the typical energy transferred in a single collision process is suppressed by a factor of order $\sim\mathcal{O}(\sqrt{\lambda}T_{\text{dS}}/m)\ll1$ in comparison to the usual expectation from thermal field theories, $E\sim T$.
We further computed the full correlator $\langle v(t)v(0)\rangle$ in (\ref{vv}) and checked that in the limit $t\to0$ we recover the previous result for $\langle v^2\rangle$. One might be tempted to compare (\ref{vv}) with the results of \cite{Oshita:2014dha} at weak coupling. However, note that the leading term reported in that paper presents a logarithmic divergence that exists even in the $H\to0$ limit. This divergence might be understood as coming from the short-distance motion of the particle, and originates due to the fact that in that formalism the particle is effectively point-like. As explained in section \ref{hbmsec}, the Brownian particle in our setup is automatically `dressed' or `composite', with a gluonic cloud of size $\sim \sqrt{\lambda}/m$, and therefore we do not expect this divergence. In order to understand the effects of this non-abelian dressing, it would be interesting to compute, at least for a static quark, the backreacted profiles of observables such as $\langle T_{\mu\nu}(r)\rangle$ and $\langle\mathrm{Tr}\,F^2(r)\rangle$. Given that the zero mass limit of the hyperbolic black hole considered here is isometric to AdS, the graviton and dilaton propagators are the same as in pure AdS (under an appropriate coordinate transformation) and the computations should proceed as in \cite{Hovdebo:2005hm,Chernicoff:2011vn,Agon:2014rda}. Finally, note also that in the large mass regime of \cite{Oshita:2014dha}, equipartition does not work either, which is consistent with our results.

On the other hand, in the late time regime the particle undergoes a regime of slow diffusion which, as explained in section \ref{dssec}, is consistent with the theory of Brownian motion in an inhomogeneous medium. In the strict limit $t\to\infty$, in particular, the displacement squared approaches a constant value inside the horizon (\ref{latetimes2}). Intuitively, the fact that the particle does not reach the
horizon is due to the fact that the frequency of interactions with Hawking quanta increases substantially as it moves away from the origin. At the equilibrium position then, the random kicks received from the thermal bath counteract the gravitational repulsion due to the de Sitter geometry. This is an example where a classical effect can be canceled out due to quantum mechanical effects, which are enhanced in the strong coupling regime. See \cite{Hu:1993qa} for a relevant discussion on this point. We also showed that the motion of the Brownian particle is in general slower than that of a test particle (that does not interact with Hawking radiation) on a radial time-like geodesic, which does reach the horizon in the $t\to\infty$ limit. One might wonder if, at strong coupling, a similar effect appears for charged particles in a black hole geometry and could there possibly be, at weak coupling, a remnant effect.

Another interesting result of this paper is the computation of the random force correlator $\kappa(\omega)$ (\ref{kappaomega}) and the friction kernel $\gamma(\omega)$ (\ref{gammaomega}). These are functions that characterize the generalized Langevin equation (\ref{genLE}), which in turn is found to be consistent with Brownian motion in an inverted harmonic potential--- see (\ref{koverm}) and the discussion below. In the low frequency limit, in particular, we obtained the expected result for thermal field theories in $1+1$ dimensions, (\ref{kappa0}) and (\ref{gamma0}), respectively. The fact that, for de Sitter space, $\kappa_0$ and $\gamma_0$ turn out to be independent of the number of
dimensions $d$ is a surprising result of the AdS/CFT correspondence. On one hand, the thermodynamics of the hyperbolic black holes considered here do depend on $d$, as shown in \cite{Emparan:1998he,Birmingham:1998nr,Emparan:1999gf}. Furthermore, the holographic stress-energy tensor, computed in \cite{Marolf:2010tg,Fischler:2013fba}, is also found to depend on $d$. We believe that this behavior is due to the fact that the function $f(z)$ (\ref{fdez}) appearing in the metric (\ref{wsmetric2}) is independent of the number of dimensions, whereas for a Schwarzschild-AdS$_{d+1}$ black hole the power of the $z$-term depends explicitly on $d$ \cite{brownian}. Moreover, the zero mass limit of the hyperbolic black holes is diffeomorphic to AdS, which is only true for the BTZ (or Schwarzschild-AdS$_{2+1}$) black hole. Aside from this peculiarity, we found that $\kappa_0$ and $\gamma_0$ satisfy the fluctuation dissipation theorem (\ref{flucdis}). More generally, the full functions $\kappa(\omega)$ and $\gamma(\omega)$ are found to be related through (\ref{generalFG}), which is known as the second fluctuation dissipation theorem--- see appendix \ref{subsub3} for details. Finally, in section \ref{remarks} we computed the real-time coefficients, $\kappa(t)$ and $\gamma(t)$, and we commented on the relevant time scales that characterize the evolution of the Brownian particle. These time scales are found to agree with the results for thermal field theories at strong coupling \cite{Atmaja:2010uu}.

A number of possibilities for the extension of our analysis can be considered. The most straightforward one is to consider other states of the CFT like the ones studied in \cite{Marolf:2010tg,Fischler:2013fba} or a general $\alpha$-vacuum. Beside this, one could for instance study the drag force exerted on a heavy quark away from the non-relativistic limit. This could be done by constructing the standard trailing string configuration as in \cite{Herzog:2006gh,Gubser:2006bz}. However, in de Sitter space we do not have invariance under spatial translations and, therefore, we do not expect such stationary solutions to exist. This could be circumvented by considering a quark moving along a circle of radius $R<1/H$, following the work of \cite{Fadafan:2008bq}. This configuration is interesting on its own right: we can obtain information about the energy loss due to collisions and, in addition, its interplay with the energy loss via synchrotron radiation which exists even in the $H\to0$ limit. We could also consider longitudinal and transverse fluctuations of the rotating solution as in \cite{gubserqhat,ctqhat} and then determine the jet-quenching parameter $\hat{q}$ of the de Sitter medium. Another interesting possibility is to consider other kinds of probes. For instance, we could study Wilson loops by looking at string configurations with the two endpoints anchored to the boundary \cite{Maldacena:1998im,Rey:1998bq,Brandhuber:1998bs}. From this, we can extract the quark-antiquark potential $V_{q\bar{q}}$ and the screening length $\ell_{\text{scr}}$ of the de Sitter medium. We are currently doing some research along these lines \cite{progress}. Going a bit further, we could also study rotating configurations as in \cite{Sadeghi:2010zzb} in order to analyze the effects of the angular momentum on the screening length. Furthermore, the light-like limit of these configurations would be a useful alternative for the analysis of jet-quenching in de Sitter space \cite{Liu:2006ug}. Finally, as mentioned in the introduction, we could study energy-momentum correlators and extract from these the various transport coefficients of the plasma.

\section*{Acknowledgements}
This material is based upon work supported by the National Science Foundation under Grant Number PHY-1316033
and by Texas Cosmology Center, which is supported
by the College of Natural Sciences and the Department of Astronomy at the University
of Texas at Austin and the McDonald Observatory.

\appendix
\section{Radial geodesics in the static patch\label{geoapp}}
In the static patch of de Sitter (\ref{staticdS}), the Lagrangian for time-like geodesics is:
\begin{equation}
L = -\frac{m}{2}g_{\mu\nu}U^{\mu}U^{\nu} = \frac{m}{2}\left[(1-H^{2}r^{2})\dot{t}^{2}-\frac{\dot{r}^{2}}{1-H^{2}r^{2}}-r^{2}\dot{\theta}^{2}-r^{2}\sin^{2}{\theta}\dot{\phi}^{2}\right]\,,
\end{equation}
where the overdot denotes differentiation with respect to the proper time. Consider a particle moving radially outward. Without loss of generality, we choose $\phi=0$ and $\theta=\frac{\pi}{2}$. Since the spacetime is static, the energy is conserved
\begin{equation}\label{energy}
E = \frac{\partial L}{\partial \dot{t}} = m\dot{t}(1-H^{2}r^{2})\,.
\end{equation}
The equation of motion for $r$ reads:
\begin{equation}\label{rtau}
\ddot{r}(1-H^{2}r^{2})-H^{2}r(1-H^{2}r^{2})^{2}\dot{t}^{2} + H^{2}r\dot{r}^{2} = 0\,.
\end{equation}
Combining (\ref{energy}) and (\ref{rtau}), we obtain a differential equation for $r(\tau)$:\footnote{This can also be cast as an equation for $r(t)$, according to (\ref{radgeo}).}
\begin{equation}\label{radialgeo}
\ddot{r}(1-H^{2}r^{2})-H^{2}r\frac{E^{2}}{m^2}+H^{2}r\dot{r}^{2}=0\,.
\end{equation}
It can be checked that the following solution is radially outward and correctly normalized ($U^{\mu}U_{\mu}=-1$):
\begin{equation}
r{(\tau)} = \frac{1}{H}\cosh{(H\tau)} - \frac{E^2}{2Hm^{2}}e^{-H\tau}\,,
\end{equation}
\begin{equation}
t{(\tau)} = \frac{1}{H}\mathrm{arctanh}{\left(\frac{m^{2}e^{2H\tau}-E^{2}-m^{2}}{2Em}\right)} - C\,,
\end{equation}
where $C$ is a constant of integration. Eliminating the proper time, we find:
\begin{equation}
r{(t)} = \frac{1}{H}\left(\frac{m+E\tanh{[H(t+C)]}}{\sqrt{m^{2}+E^{2}+2Em\tanh{[H(t+C)]}}}\right)\,.
\end{equation}
In particular, $r \rightarrow 1/H$ as $t \rightarrow \infty$. We will now choose $C$ as
\begin{equation}
C = -\frac{1}{H}\mathrm{arctanh}{\left(\frac{m}{E}\right)}\,,
\end{equation}
so that $r(0)=0$. Substituting back into the solution, we finally obtain:
\begin{equation}
r(t) = \frac{1}{H}\left[\frac{m+E\tanh{(Ht-\mathrm{arctanh}(m/E))}}{\sqrt{m^{2}+E^{2}+2Em\tanh{(Ht-\mathrm{arctanh}(m/E))}}}\right]\,.
\end{equation}
Using the special relativistic relation $E = \gamma_0\, m$, where $\gamma_0=(1-v_{0}^2)^{-1/2}$, we can rewrite the solution in terms of the initial velocity $v_{0}$:
\begin{equation}
r(t) = \frac{1}{H}\left[ \frac{\sqrt{1-v_{0}^{2}}+\tanh{(Ht-\mathrm{arctanh}{\sqrt{1-v_{0}^{2}}})}}{\sqrt{2-v_{0}^{2}+2\sqrt{1-v_{0}^{2}}\tanh{(Ht-\mathrm{arctanh}{(\sqrt{1-v_{0}^{2}})})}}}\right]\,.
\end{equation}
In particular, in the non-relativistic limit, we find the expected result at early times
\begin{equation}\label{earlytimer}
r(t) \approx v_{0}t + \frac{1}{6}H^{2}v_{0}t^{3}\,,
\end{equation}
where the lowest order term correspond to inertial motion, and the next term describes acceleration in an inverted harmonic potential of the form (\ref{pot}).

\section{Normalization of the mode functions\label{normalizationSEC}}

In this appendix we will normalize the mode functions (\ref{modefnc}) with respect to the Klein-Gordon inner product \cite{bd}
\begin{equation}\label{kgip}
(f_{i},g_{j})_{\sigma} = -i\frac{1}{2\pi \alpha'} \int_{\sigma} \sqrt{\tilde{g}} n^{\alpha} G_{ij} (f_{i}\partial_{\alpha}g_{j}^{*}-\partial_{\alpha}f_{i}g_{j}^{*})\,.
\end{equation}
Here $\sigma$ is a Cauchy surface in the worldsheet metric (\ref{wsmetric2}), $\tilde{g}$ is the induced metric on $\sigma$ and $n^{\alpha}$ is the unit normal vector to $\sigma$. For simplicity, we take $\sigma$ to be a constant-$t$ surface. In this case, the determinant of the induced metric is found to be
\begin{equation}
\sqrt{\tilde{g}} = \sqrt{g_{zz}}=\frac{L}{z}\,.
\end{equation}
Moreover, the normal vector $n^{\alpha}$ will have the form $n^{t} = F(t,z)\partial_{t}$, where $F$ is a function such that $g_{tt}F^{2} = -1$. From this it follows that $F = z/L f(z)$, and the Klein-Gordon inner product becomes
\begin{equation}\label{KG}
(u_{\omega},u_{\omega'}) = -i\frac{\sqrt{\lambda}}{2\pi} \int  dz\frac{f(z)}{z^2}(u_{\omega}\dot{u}_{\omega'}^{*}-\dot{u}_{\omega}u_{\omega'}^{*})\,,
\end{equation}
where
\begin{equation}
u_{\omega}\dot{u}_{\omega'}^{*}-\dot{u}_{\omega}u_{\omega'}^{*} = -iA_{\omega}A_{\omega'}(\omega+\omega')e^{i(\omega'-\omega)t}(g_{\omega}^{(\text{in})}+B_{\omega}g_{\omega}^{(\text{out})})(g_{\omega'}^{(\text{in})*}+B_{\omega'}^{*}g_{\omega'}^{(\text{out})*})\,.
\end{equation}

We will now normalize the modes such that $(u_{\omega},u_{\omega'})=2\pi\delta{(\omega-\omega')}$. Of course, there is a contribution to the inner product coming from all regions of spacetime, but because the near horizon region is semi-infinite, the normalization of solutions is completely determined by this region \cite{Atmaja:2010uu}. In the near horizon limit (\ref{KG}) becomes:
\begin{equation}\label{nearhorizonKG}
(u_{\omega},u_{\omega'}) = i\frac{\sqrt{\lambda}H^{2}}{8\pi} \int dz_{*} f(z_*)^2 (u_{\omega}\dot{u}_{\omega'}^{*}-\dot{u}_{\omega}u_{\omega'}^{*})\,,
\end{equation}
where $z_*\in(-\infty,0]$ is the tortoise coordinate defined in (\ref{tortoisez}) and
\bea
&&g_{\omega}^{(\text{in})} = \frac{2}{f(z_*)}\left(1-i\frac{\omega}{H}\right)e^{-i\omega z_{*}}\,,\label{inzst}\\
&&g_{\omega}^{(\text{out})} = g_{\omega}^{(\text{in})*}\,.\label{outzst}
\eea
Substituting (\ref{inzst}) and (\ref{outzst}) into (\ref{nearhorizonKG}), we find
\begin{equation}
(u_{\omega},u_{\omega'}) = \frac{\sqrt{\lambda}H^{2}}{8\pi}A_{\omega}A_{\omega'}^{*}(\omega+\omega')e^{i(\omega'-\omega)t}\left[J_{1}+J_{2}+J_{3}+J_{4}\right]\,,
\end{equation}
where
\bea
&&J_{1} = \left(2-2i\frac{\omega}{H}\right)\left(2+2i\frac{\omega'}{H}\right)\int_{-\infty}^{0} dz_{*} e^{-i(\omega-\omega')z_{*}}\,,\nonumber\\
&&J_{2} = B_{\omega}B_{\omega'}^{*}\left(2+2i\frac{\omega}{H}\right)\left(2-2i\frac{\omega'}{H}\right)\int_{-\infty}^{0} dz_{*} e^{i(\omega-\omega')z_{*}}\,,\nonumber\\
&&J_{3} = B_{\omega'}^{*}\left(2-2i\frac{\omega}{H}\right)\left(2-2i\frac{\omega'}{H}\right)\int_{-\infty}^{0} dz_{*} e^{-i(\omega+\omega')z_{*}}\,,\nonumber\\
&&J_{4} = B_{\omega}\left(2+2i\frac{\omega}{H}\right)\left(2+2i\frac{\omega'}{H}\right)\int_{-\infty}^{0} dz_{*} e^{i(\omega+\omega')z_{*}}\,.\nonumber
\eea
To evaluate these integrals, we use the formula\footnote{The term $\frac{i}{a}$ in this formula should be understood as the distribution $\mathcal{P}\left(\frac{i}{a}\right)$ where $\mathcal{P}$ denotes the Cauchy principal value.}
\begin{equation}
\int_{-\infty}^{0} e^{i a x} dx = \pi \delta{(a)} - \frac{i}{a}\,.
\end{equation}
We then have:
\bea
&&J_{1} = 4\pi\left(1+\frac{\omega^{2}}{H^{2}}\right)\delta{(\omega-\omega')} + \frac{i}{\omega-\omega'}\left(2-2i\frac{\omega}{H}\right)\left(2+2i\frac{\omega'}{H}\right)\,,\nonumber\\
&&J_{2} = 4\pi\left(1+\frac{\omega^{2}}{H^{2}}\right)\delta{(\omega-\omega')} - \frac{i}{\omega-\omega'}\left(2-2i\frac{\omega'}{H}\right)\left(2+2i\frac{\omega}{H}\right)\,,\nonumber\\
&&J_{3} = 4\pi B_{\omega}\left(1+\frac{\omega^{2}}{H^{2}}\right)\delta{(\omega+\omega')} + \frac{i}{\omega+\omega'}\left(2-2i\frac{\omega}{H}\right)\left(2-2i\frac{\omega'}{H}\right)B_{\omega}\,,\nonumber\\
&&J_{4} = 4\pi B_{\omega}\left(1+\frac{\omega^{2}}{H^{2}}\right)\delta{(\omega+\omega')} - \frac{i}{\omega+\omega'}\left(2+2i\frac{\omega}{H}\right)\left(2+2i\frac{\omega'}{H}\right)B_{\omega}\,,\nonumber
\eea
where we used the fact that $B_{\omega}B_{\omega}^{*}=1$ and $B_{-\omega}=B_{\omega}^{*}$. Adding these four terms up we get
\begin{equation}
J_{1}+J_{2}+J_{3}+J_{4} = 8\pi\left(1+\frac{\omega^{2}}{H^{2}}\right)\big[\delta{(\omega-\omega')} + B_{\omega} \delta{(\omega+\omega')}\big]\,.
\end{equation}
We argue that the term proportional to $\delta{(\omega+\omega')}$ does not contribute. Since both $\omega$ and $\omega'$ are positive frequencies, this is non-zero only when $\omega=\omega'=0$ but in that case, the overall factor in front of $(u_{\omega},u_{\omega'})$ vanishes. Thus, at the end we are left with
\begin{equation}
(u_{\omega},u_{\omega'}) = 2\sqrt{\lambda}|A_{\omega}|^{2} \omega\left(H^{2}+\omega^{2}\right)\delta{(\omega-\omega')}=2\pi\delta{(\omega-\omega')}\,,
\end{equation}
and solving for $A_{\omega}$ we obtain (\ref{constantA}). Following the same lines, we can easily check that $(u^*_{\omega},u^*_{\omega'})=-(u_{\omega},u_{\omega'})$ and $(u_{\omega},u^*_{\omega'})=0$. This ensures that the canonical commutation relations are satisfied:
\begin{align}
  [a_{\omega},a_{\omega'}]=
  [a_{\omega}^\dagger,a_{\omega' }^\dagger]=0,\qquad
  [a_{\omega},a_{\omega'}^\dagger]=2\pi\delta(\omega-\omega').\label{CCR}
\end{align}

\section{Evaluation of the displacement squared\label{s2app}}
In this appendix we show some of the steps needed for the evaluation of the integral appearing in (\ref{s2}), which has the following form:
\begin{equation}
I = \int_{0}^{\infty} dx \frac{x}{1+a^{2}x^{2}} \frac{\sin^{2}(\frac{b x}{2})}{e^{x}-1}\,.
\end{equation}
Here $a = 1/2\pi$ and $b = Ht/2\pi$. To do this, we consider a similar integral:
\begin{equation}\label{I1int}
I_{1} = \int_{0}^{\infty} dx \frac{1}{x(1+a^{2}x^{2})} \frac{\sin^{2}(\frac{b x}{2})}{e^{x}-1}\,,
\end{equation}
which was computed in \cite{brownian} using contour integration methods. Using
\begin{equation}
\frac{\partial^{2}}{\partial b^{2}} \sin^{2}{\left(\tfrac{b x}{2}\right)} = \frac{x^2}{2}-x^{2}\sin^{2}{\left(\tfrac{b x}{2}\right)}\,,
\end{equation}
we find
\begin{equation}
I= I_{2}-\frac{\partial^{2}}{\partial b^{2}}I_{1}\,,
\end{equation}
where $I_{2}$ is the integral
\begin{equation}\label{I2appC}
I_{2} = \frac{1}{2} \int_{0}^{\infty} dx \left(\frac{x}{1+a^{2}x^{2}}\right)\left(\frac{1}{e^{x}-1}\right)= -\frac{1}{4a^{2}}\left(a\pi+\log{(2a\pi)}+\psi{\left(\tfrac{1}{2a\pi}\right)}\right)\,.
\end{equation}
The function $\psi(z)=(d/ dz)\log\Gamma(z)$ appearing above is the digamma function. For $a=1/2\pi$, in particular, this reduces to
\begin{equation}
I_{2} = -\left(\tfrac{1}{2}-\gamma\right)\pi^{2}\,.
\end{equation}
The integral $I_1$ evaluates to \cite{brownian}
\begin{eqnarray}
&&I_{1} = \frac{1}{8}\left[e^{b/a}\mathrm{Ei}{\left(-\tfrac{b}{a}\right)}+e^{-b/a}\mathrm{Ei}{\left(\tfrac{b}{a}\right)}\right]+\frac{1}{8}\left[\psi{(1+\tfrac{1}{2\pi a})}+\psi{(1-\tfrac{1}{2\pi a})}\right] \nonumber \\
&&\qquad\quad+ \frac{e^{-2\pi|b|}}{8}\left[\frac{{}_{2}F_{1}{(1,1+\frac{1}{2\pi a},2+\frac{1}{2\pi a},e^{-2\pi|b|})}}{1+\frac{1}{2\pi a}}+\frac{{}_{2}F_{1}{(1,1-\frac{1}{2\pi a},2-\frac{1}{2\pi a},e^{-2\pi|b|})}}{1-\frac{1}{2\pi a}}\right] \nonumber \\
&&\qquad\quad-\frac{\pi}{8}(1-e^{-|b|/a})\cot{\left(\tfrac{1}{2a}\right)} + \frac{1}{4}\log{\left(\tfrac{2a}{b}\sinh{(\pi b)}\right)}\,,
\end{eqnarray}
where ${\rm Ei}(z)$ is the exponential integral and
${}_2F_1(\alpha,\beta,\gamma,z)$ is the hypergeometric function. For
${\rm Ei}(z)$ we take the branch where both ${\rm Ei}(x>0)$ and
${\rm Ei}(x<0)$ are real. This expression looks singular at $a=1/2\pi$ but in fact it is not. Taking the second derivative with respect to $b$ yields:
\begin{eqnarray}
&&\frac{\partial^{2}}{\partial {b}^{2}}I_{1} = \frac{1}{8a^{2}}\left[e^{b/a}\mathrm{Ei}{\left(-\tfrac{b}{a}\right)}+e^{-b/a}\mathrm{Ei}{\left(\tfrac{b}{a}\right)} + \pi e^{-b/a}\cot{(\frac{1}{2a})}\right] \\
&&\qquad\qquad+ \frac{e^{-2\pi b}}{8a^{2}}\left[\frac{{}_{2}F_{1}{(1,1-\frac{1}{2\pi a},2-\frac{1}{2\pi a},e^{-2\pi b})}}{1-1/2\pi a} + \frac{{}_{2}F_{1}{(1,1+\frac{1}{2\pi a},2+\frac{1}{2\pi a},e^{-2\pi b})}}{1+1/2\pi a}\right]\,.\nonumber
\end{eqnarray}
Next, we take the limit $a \rightarrow 1/2\pi$. Two of the terms in the expression above diverge in this limit, but we find that their divergences cancel each other. We need to compute the following:
\begin{equation}
L = \lim_{a \rightarrow 1/2\pi} \left[\pi e^{-b/a}\cot{(1/2a)} + e^{-2\pi b}\frac{{}_{2}F_{1}{(1,1-1/2\pi a,2-1/2\pi a,e^{-2\pi b})}}{1-1/2\pi a} \right]\,.
\end{equation}
To avoid cumbersome expressions, we will define the parameter $\alpha = 1 - 1/2\pi a$.
Expanding the first term in series around $\alpha=0$ and using the series representation of the hypergeometric function, the limit above becomes:
\begin{equation}
L = \lim_{\alpha \rightarrow 0} \left[-\frac{e^{-2\pi b}}{\alpha}-2\pi be^{-2\pi b} + \mathcal{O}(\alpha) + e^{-2\pi b} \sum_{n=0}^{\infty} \frac{e^{-2\pi b n}}{n+\alpha} \right]\,.
\end{equation}
Finally, taking the limit $\alpha \rightarrow 0$, and recognizing in the remaining terms as the Taylor series of the logarithm, we find:
\begin{equation}
L = -e^{-2\pi b}(2\pi b + \log{(1-e^{-2\pi b})})\,.
\end{equation}
Putting everything together, and expressing the final result as a function of $t$, we find the expression for the displacement squared reported in (\ref{finals2}).

\section{The distribution of momentum\label{subsub4}}

In this appendix, we derive the probability distribution $f(p)$ for the momentum of the Brownian particle. For this purpose, it is convenient to first revisit the boundary conditions for the field $X$, specially in the IR of the geometry. While it is true that the normalization of the mode functions via the Klein-Gordon inner product is fixed by the near horizon region, a close inspection of (\ref{kgip}) reveals that it suffers from an IR divergence \cite{brownian}. To regularize this divergence, we have to impose another cutoff near the horizon. To this effect, we cut off the geometry in the IR at the stretched horizon $z_s\equiv\frac{2}{H}(1-2\epsilon)$. Now, imposing a Neumann boundary condition\footnote{In the $\epsilon\to0$ limit we could alternatively take a Dirichlet boundary condition, but this would make no difference.} at $z_s$, we get an extra condition on $B_{\omega}$:
\begin{equation}
B_\omega = -e^{2i\frac{\omega}{H}\ln{(1/\epsilon)}}\,,
\end{equation}
which discretizes the allowed values for $\omega$. Furthermore, the overall normalization $A_\omega$ is also modified by the introduction of this IR cutoff. After some algebra, we get
\be
A_{\omega} = \left(\frac{\pi H}{2 \sqrt{\lambda}\omega(H^{2}+\omega^{2})\ln{(1/\epsilon)}}\right)^{\!\!\frac{1}{2}}\,.
\ee

All mode decompositions become summations over the discrete set of allowed frequencies.\footnote{However, the spacing of frequencies is small in the $\epsilon\to0$ limit and these summations can be well approximated by integrals, taking into account the density of states
$$
\sum_{\omega\geq0} \longrightarrow \int_0^{\infty} d\omega \mathcal{D}(\omega)\,,\qquad \mathcal{D}(\omega) \equiv \frac{1}{\Delta \omega}= \frac{\ln{(1/\epsilon)}}{\pi H}\,.
$$
This step has been implicit in all of our computations.}
In particular, the momentum operator becomes
\begin{equation}
p{(t)} = -2 m z_m \sum_{\omega \geq 0} \left[a_{\omega}\omega^{2}A_{\omega}\left(\frac{2+Hz_{m}}{2-Hz_{m}}\right)^{i\frac{\omega}{H}}e^{-i\omega t} + \mathrm{h.c.}\right]\,.
\end{equation}
In order to obtain the distribution function $f(p)$ note that, by definition:
\begin{equation}\label{eixip}
\langle e^{i p \xi} \rangle = \int_{-\infty}^{\infty} dp e^{ip\xi} f(p)\,.
\end{equation}
Given the above quantity, then, we can simply apply the inverse Fourier transform to obtain $f(p)$. Using normal ordering, the left-hand side of (\ref{eixip}) becomes
\begin{equation}\label{eipxi}
\langle :\!e^{ip\xi}\!:\rangle = \langle:\! \exp{\left(\sum_{\omega \geq 0} a_{\omega}\alpha_{\omega} - \alpha_{\omega}^{*}a_{\omega}^{\dagger}\right)} \!:\rangle = \frac{1}{Z} \mathrm{Tr} \left[e^{-\beta \mathcal{H}} :\! \exp{\left(\sum_{\omega \geq 0} a_{\omega}\alpha_{\omega} - \alpha_{\omega}^{*}a_{\omega}^{\dagger}\right)} \!:\right]\,,
\end{equation}
where
\begin{equation}
\alpha_{\omega} = -2im z_{m}  \omega^{2}A_{\omega}\xi\left(\frac{2+Hz_{m}}{2-Hz_{m}}\right)^{i\frac{\omega}{H}}e^{-i\omega t}\,,
\end{equation}
and $\mathcal{H}$ is the thermal Hamiltonian
\begin{equation}
\mathcal{H} = \sum_{\omega \geq 0} \omega a_{\omega}^{\dagger} a_{\omega}\,.
\end{equation}
Using the fact that ladder operators of different frequencies commute, we can rewrite (\ref{eipxi}) as
\begin{equation}
\langle :\!e^{ip\xi}\!:\rangle = \frac{1}{Z}\mathrm{Tr} \prod_{\omega \geq 0} {\left(e^{-\beta\omega a_{\omega}^{\dagger}a_{\omega}} :\!e^{a_{\omega}\alpha_{\omega}-\alpha_{\omega}^{*}a_{\omega}^{\dagger}}\!: \right)}\,.
\end{equation}
Now, we use the formula
\begin{equation}
\mathrm{Tr}{\left(e^{-\beta\omega a^{\dagger}a}:\!e^{\alpha a - \alpha^{*}a^{\dagger}}\!:\right)} = \frac{1}{1-e^{-\beta\omega}} \exp{\left(-\frac{|\alpha|^{2}}{e^{\beta\omega}-1}\right)}
\end{equation}
to obtain
\begin{equation}
\langle:\! e^{ip\xi} \!:\rangle = C \exp{\left(-\sum_{\omega \geq 0} \frac{|\alpha_{\omega}|^{2}}{e^{\beta\omega}-1}\right)}\,,
\end{equation}
where $C$ is a constant independent of $\xi$. Substituting for $\alpha_{\omega}$ and $A_{\omega}$ and replacing the sum by an integral, we find
\begin{equation}
\langle:\! e^{ip\xi} \!:\rangle = C \exp{\left(-\frac{\xi^{2}\sqrt{\lambda}H^{2}}{32\pi^{6}} \int_{0}^{\infty} \frac{dx}{e^{x}-1} \frac{x^3}{1+\frac{x^2}{4\pi^{2}}}\right) }\,.
\end{equation}
We will now explicitly evaluate the integral above. To do this, notice that
\begin{equation}
 \int_{0}^{\infty} \frac{dx}{e^{x}-1} \frac{x^3}{1+\frac{x^2}{4\pi^{2}}}=-2\frac{\partial^{4}I_{1}}{\partial b^{4}}\bigg|_{b=0}\,,
\end{equation}
where $I_{1}$ is the integral (\ref{I1int}) appearing in appendix \ref{s2app}. We then find
\begin{equation}
\int_{0}^{\infty} \frac{dx}{e^{x}-1}\frac{x^3}{1+\frac{x^2}{4\pi^{2}}} = \frac{2}{3}(7-12\gamma_{E})\pi^{4}\,,
\end{equation}
and therefore
\begin{equation}
\langle:\! e^{ip\xi} \!:\rangle = C \exp{\left(-\frac{\xi^{2}\sqrt{\lambda}H^{2}(7-12\gamma_{E})}{48\pi^{2}}\right)}\,.
\end{equation}
After Fourier transforming, we find that the momentum distribution $f(p)$ is a Gaussian function, as expected by invariance under time translations, but the width is not the same as the Maxwell-Boltzmann distribution:
\begin{equation}\label{disfp}
f{(p)} \propto \exp{\left(-\frac{12\pi^{2}p^{2}}{\sqrt{\lambda}H^{2}(7-12\gamma_{E})}\right)}\,.
\end{equation}
From this distribution, we can compute the average $\langle v^{2} \rangle=\langle v(t)v(t) \rangle=\langle v(0)v(0) \rangle$:
\begin{equation}\label{vv0}
\langle v^{2} \rangle = \left(\frac{\sqrt{\lambda}H^{2}}{4\pi^{2}m^{2}}\right)\left(\frac{7}{6}-2\gamma_{E}\right)\,.
\end{equation}
This is equivalent to the quantity $v_0^2$ computed in (\ref{v0s2}). In particular, comparing the above with the early time expansion for the displacement square (\ref{ballistic}), we find that the latter can be expressed as $\langle s^{2}\rangle \approx \langle v^{2} \rangle t^{2}$, corroborating the fact that the equipartition theorem does not hold. More in general, from (\ref{ppcor}) we can compute the quantity $\langle v(t)v(0) \rangle$. Following the same steps as in appendix \ref{s2app} we can explicitly evaluate the integral appearing in (\ref{ppcor}). At the end we obtain
\bea\label{vv}
&&\!\!\langle :\!v(t)v(0)\!: \rangle = \frac{H^{2}}{4\pi^{2}\sqrt{\lambda}m^2} \bigg[ \frac{2}{H^{2}t^{2}} - \frac{2\cosh{(Ht)}}{(1-e^{Ht})^2} - \frac{2\cosh{(Ht)}}{e^{Ht}-1}+ Hte^{-Ht}  + 2 \nonumber \\
&&\!\!\qquad\qquad\qquad\quad- e^{Ht}\mathrm{Ei}{(-Ht)} - e^{-Ht}\mathrm{Ei}{(Ht)} + 2\cosh{(Ht)}\log{(1-e^{-Ht})} \bigg].
\eea
This looks singular at early times but, in fact, it reduces to (\ref{vv0}) in the limit $t\to0$. On the other hand, $\langle :\!v(t)v(0)\!: \rangle\to0$ as $t\to\infty$, meaning that the velocities at late and initial times are totally uncorrelated. The fact that (\ref{vv}) vanishes at late times, and that $\langle s^{2}\rangle$ goes to a constant value (\ref{latetimes2}), suggests that in such regime a form of the equipartition should be valid
\be
\frac{1}{2}m \langle v^{2} \rangle+ \phi_{\text{eff}}\sim T\,,
\ee
where $\phi_{\text{eff}}$ is an effective potential energy that accounts for gravitational and thermal interactions. Then, close to the equilibrium position, $\bar{x}$, we expect that
\be
\phi_{\text{eff}}\sim\phi_0+\frac{1}{2}m\bar{\omega}^2 (x-\bar{x})^2+...\,,
\ee
for some $\bar{\omega}^2>0$ and $\phi_0\sim T$. However, in order to prove such conjecture we would need further input that goes beyond the scope of this paper.

\section{The fluctuation-dissipation theorem\label{subsub3}}

In this appendix we will reconsider the random force correlator computed in \ref{subsub2} in light of a different regularization scheme. More specifically, following \cite{brownian} we will regularize by means of the canonical correlator (instead of normal-ordering), and we will show that the second fluctuation-dissipation theorem (\ref{generalFG}) holds in our holographic setup.

At finite temperature, the canonical correlator is defined by \cite{Kubo:f-d_thm}
\begin{equation}
\langle X;Y \rangle = \frac{1}{\beta} \int_{0}^{\beta} d\lambda \langle e^{\lambda H} X e^{-\lambda H} Y \rangle\,,
\end{equation}
and satisfies the following property:
\begin{equation}\label{prop}
\langle [A{(0)},B{(t)}] \rangle = i\beta \langle \dot{A}{(0)};B{(t)}\rangle\,.
\end{equation}
To compute $\langle x{(0)};x{(t)}\rangle$, we first define $y{(t)}$ to be the antiderivative of $x{(t)}$,
\begin{equation}
y{(t)} = \int x{(t)}dt = 2z_m\int_{0}^{\infty} \frac{d\omega}{2\pi} \left[a_{\omega}A_{\omega}\left(\frac{2+Hz_{m}}{2-Hz_{m}}\right)^{i\omega/H}e^{-i\omega t} + \mathrm{h.c.}\right]\,,
\end{equation}
and, from (\ref{prop}), it follows that
\begin{equation}\label{prop2}
\langle x{(0)};x{(t)}\rangle = \frac{1}{i\beta} \langle [y{(0)},x{(t)}] \rangle\,.
\end{equation}
The right-hand side of (\ref{prop2}) can be evaluated by making use of the canonical commutation relations:
\begin{equation}
[a_{\omega},a^{\dagger}_{\omega'}] = 2\pi\delta{(\omega-\omega')}\,,\qquad [a^{\dagger}_{\omega},a^{\dagger}_{\omega'}]=[a_{\omega},a_{\omega'}]=0\,.
\end{equation}
We then find:
\begin{equation}
\langle x{(0)};x{(t)}\rangle = \frac{\sqrt{\lambda}H}{\pi^2m^2} \int_{0}^{\infty} \frac{d\omega}{2\pi} \cos{(\omega t)}\frac{1}{H^{2}+\omega^{2}}\,,
\end{equation}
and
\begin{equation}\label{ppc}
\langle p{(0)};p{(t)}\rangle = \frac{\sqrt{\lambda}H}{\pi^2} \int_{0}^{\infty} \frac{d\omega}{2\pi} \cos{(\omega t)} \frac{\omega^{2}}{H^{2}+\omega^{2}}\,.
\end{equation}
From (\ref{ppc}) we obtain\footnote{The displacement squared $\langle s^2(t)\rangle$ is also modified. However, the canonical correlator does not change the early and late time limits of this quantity \cite{brownian}.}
\begin{equation}\label{Ipc}
I^{c}_{p}{(\omega)} = \frac{\sqrt{\lambda}H}{2\pi^2}\left(\frac{\omega^{2}}{H^{2}+\omega^{2}}\right)\,,
\end{equation}
where the quantity $I^{c}_{p}{(\omega)}$ denotes the power spectrum computed using the canonical correlator. From the above expression, it also follows that
\be
\kappa^c(\omega)=\frac{\sqrt{\lambda}H}{2\pi^2}\left(\frac{H^2+\omega^2}{1+\omega^2z_m^2}\right)
\ee
which is related to the friction kernel (\ref{gammaomega}) through (\ref{generalFG}). This proves that the fluctuation-dissipation theorem is indeed valid for arbitrary frequencies.

\end{document}